\documentclass[bibnotes,amsmath,amssymb,aps, physrev, pre,twocolumn,showkeys, reprint]{revtex4-2}

\usepackage{graphicx}
\usepackage{dcolumn}
\usepackage{bm}
\usepackage{physics}
\usepackage[normalem]{ulem}
\usepackage{amsmath}
\usepackage{hyperref}
\usepackage{xcolor}
\begin{document}


\title{\textbf{Topological entropy of stationary three-dimensional turbulence}}%

\author{Ankan Biswas, Amal Manoharan and Ashwin Joy}
 \email{ashwin@physics.iitm.ac.in}
\affiliation{%
 Department of Physics, Indian Institute of Technology - Madras, Chennai - 600036, India.
}%

\date{\today}
\begin{abstract}
  Topological entropy serves as a viable candidate for quantifying mixing and complexity of a highly chaotic system. Particularly in turbulence, this is determined as the exponential stretching rate of a fluid material line that typically necessitates a Lagrangian description. We extend our recent work [A. Manoharan, S. Subramanian, and A. Joy, Phys. Rev. E 112, 015106] to three dimensions, and present an exact Eulerian framework to compute the topological entropy of stationary turbulent flows. The only prerequisite is a distribution of eigenvalues of the local strain-rate tensor and their decorrelation times. This can be easily obtained from a single wire probe at a fixed location, thereby eliminating the need for Lagrangian particle tracking which is formidable due to the chaotic nature of the flow. We believe that our results lend great utility in experiments targeting transport and mixing in many industrial and natural flows.
\end{abstract}

\maketitle


\section{\label{sec:level1} Introduction}
Turbulence is ubiquitous \textemdash from the small scales of a cup of coffee to the vast cosmic scales of stars, supernovae, and accretion disks around black holes \cite{zhou2021turbulence, wensink2012meso, canuto2009turbulence, canuto1998turbulence}. Despite its ubiquity and the volume of research carried out in the last century, turbulence remains one of the most challenging problems in classical physics that has attracted the attention of many \cite{reynolds1883xxix, kolmogorov1968local, kolmogorov1991dissipation}. Several unsolved problems exist in classical turbulence ranging from the existence and smoothness issues in the governing equations in three-dimensional (3D) flows, through deviations in Kolmogorov's energy cascade theory, to even anomalous scaling laws of transport \cite{fefferman2006existence,alexakis2018cascades,falkovich2001particles}. A major challenge in modern turbulence research is quantifying the emergent complexity of material flow lines. This is relevant as it is fundamentally related to transport and dissipation in many industrial and natural flows. In practice, this has been achieved through the Lyapunov exponent \cite{wolf1985determining, oseledec1968multiplicative, wolf1986quantifying}, fractal dimensions \cite{meneveau1987multifractal}, and the topological entropy \cite{adler1965topological, ott2002chaos, newhouse1993estimation, goodwyn1969topological, bowen1973topological, bowen1970topological}. Among these, the work of Newhouse and Pignataro \cite{newhouse1993estimation} stands out as a particularly convenient method to apply in practice: the topological entropy can be determined as the exponential stretching rate of a material line over arbitrary time. It gives an idea of the rate at which information about the initial condition is lost as the flow progresses and provides an insight to quantify the exponential divergence of distinguishable orbits in a dynamical system. Unlike the Lyapunov exponent that measures the rate at which nearby trajectories diverge, the topological entropy provides a global characterization of the flow field by taking into account large-scale structures, making it a preferred metric for analyzing large-scale mixing dynamics of oceanographic and atmospheric flows. However, the description of topological entropy involves the Lagrangian tracking of a large number of particles, which presents a formidable challenge when the flow is turbulent \cite{westerweel2013particle, toschi2009lagrangian, schanz2016shake}. Recently, we eliminated the need for this Lagrangian tracking by presenting an exact Eulerian framework for the topological entropy that only required a distribution of eigenvalues of the strain-rate tensor at any given location of a two-dimensional (2D) stationary and ergodic flow \cite{mdkl-nnq1}. This presents a great opportunity to experimentalists dealing with highly chaotic flows where particle trajectories are entangled and often poorly resolved. In this paper, we extend the work of \cite{mdkl-nnq1} to 3D flows, thus bringing a wide variety of flows under the ambit of our Eulerian framework. This extension is crucial, as real-world turbulent flows are inherently three-dimensional and a proper understanding requires going beyond the constraints of a 2D analysis. We believe our work should significantly augment the understanding of particle advection and mixing in many natural \cite{stohl1998computation,peacock2013lagrangian,thiffeault2010braids,Hazpra-S,haszpra2011volcanic} and industrial flows \cite{ottino1990mixing, zahtila2023particle, kusters1991influence}. The paper is organized as follows. In Section II, we provide a theory for the Eulerian framework of the topological entropy. Section III describes the numerical methods and results. Finally, in Section IV, we conclude our observations and  suggest possible directions for future research.

\begin{figure*}[t] 
\raggedright
        \includegraphics[width=\textwidth]{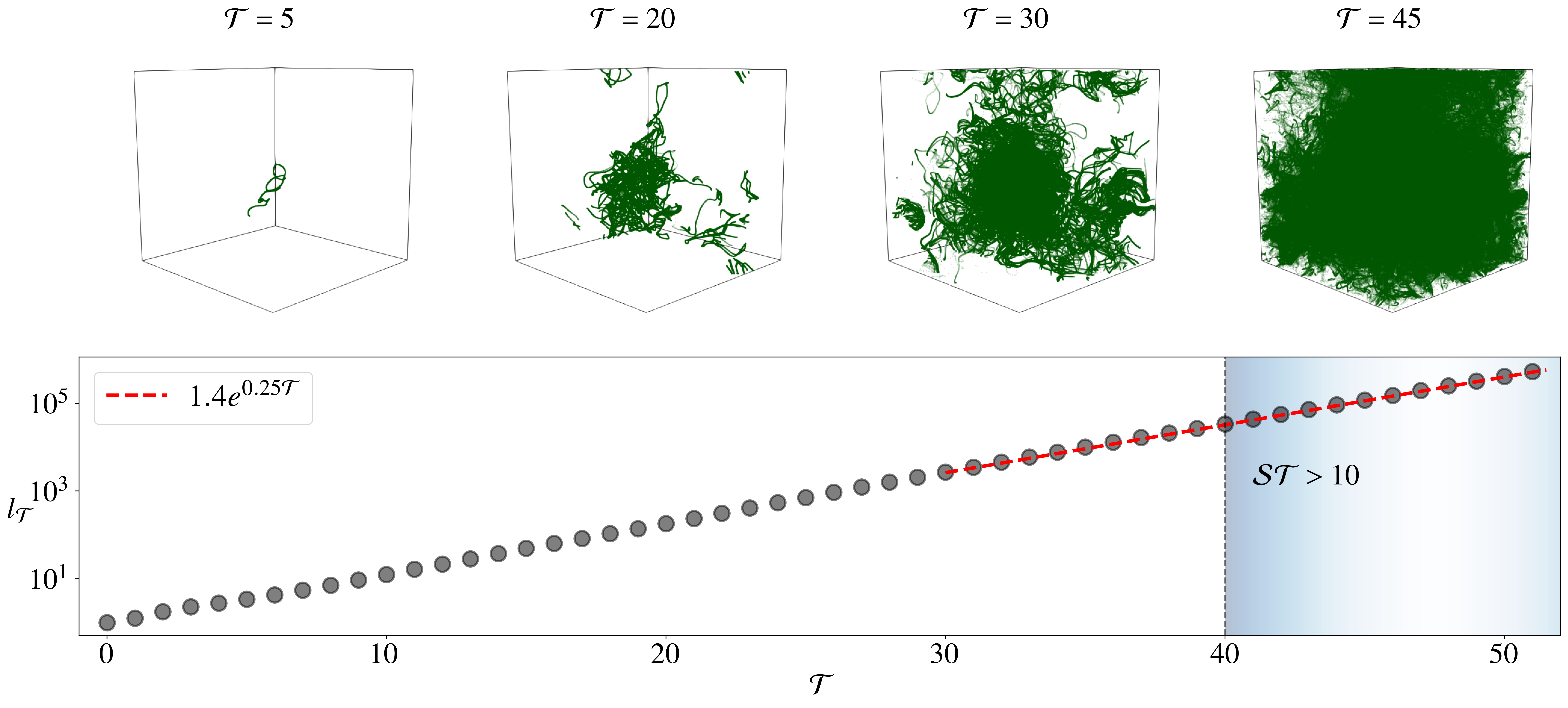}
  \caption{Top: Time evolution of the material curve initialized as a straight line in a typical numerical simulation of a turbulent flow at  Re $=10^3$. The domain of the flow is a periodic box of size $(2\pi)^3$. For the details of the numerical simulation, refer Section \ref{sec:level3}. Bottom: The exponential growth rate at late time ($\mathcal{ST} > 10$) provides the topological entropy of the flow that is plotted in Fig. \ref{fig:top_ent} as the Lagrangian measure.}
  \label{fig:material-line}
\end{figure*}

\section{\label{sec:level2} Theory}
\begin{figure*}[t!]
  \centering
          \includegraphics[width=0.9\linewidth]{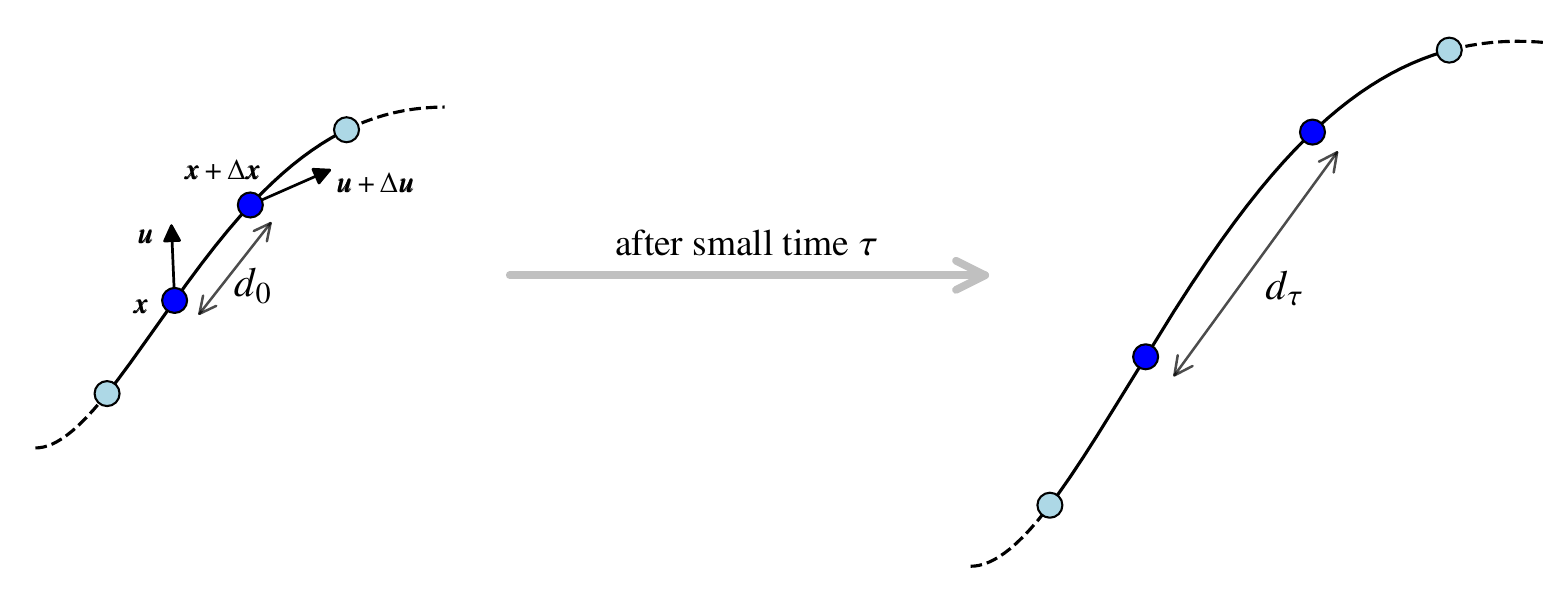}
	  \caption{A typical tracer pair separated by a distance $d_0$ deforms to $d_\tau$ over a small time $\tau$}.
      \label{fig:line_stretching}
\end{figure*}

\begin{figure}[h]
\raggedright
    \includegraphics[width=\columnwidth]{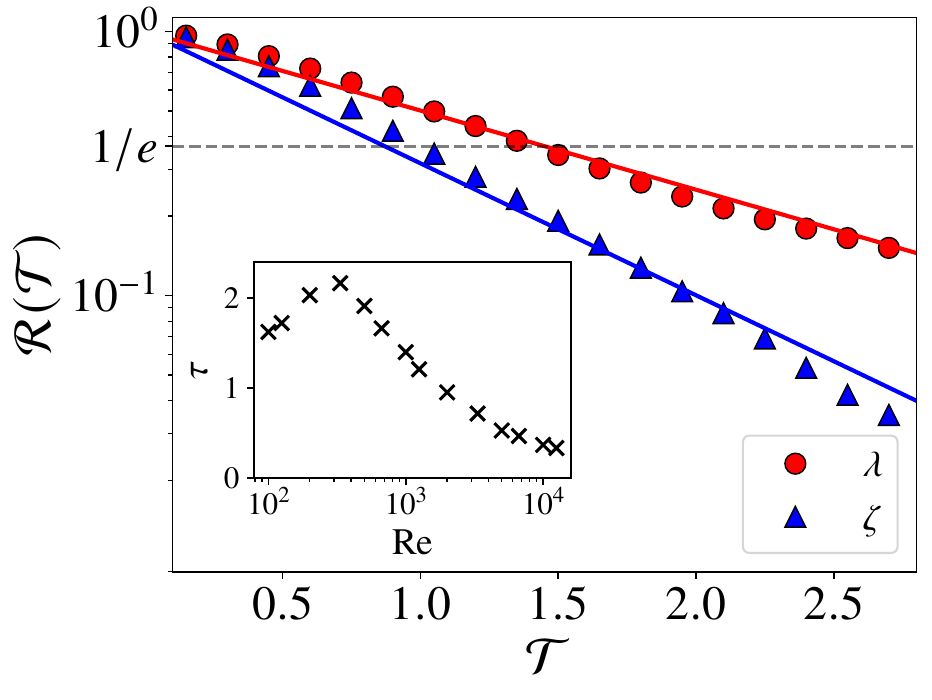}
    \caption{Auto-correlation functions of the eigenvalues $\lambda$ and $\zeta$ as defined in Eq. \ref{eqn:acf} for a representative Reynold's number Re $=10^3$. Solid lines are fits to the exponential decays. The e-folding time of the eigenvalue $\lambda$ is taken as $\tau \approx 1.5$ as it exhibits the slower decay. This is the working protocol at all Re. Inset: Decorrelation time as a function of Re. One must typically wait for a time $\mathcal{T} \gg \tau$ for the material curve to reach exponential growth.}  
    \label{fig:decorr_time}
\end{figure}

\begin{figure}[h] 
\vspace{15pt}
    \raggedright
\includegraphics[width=\linewidth]{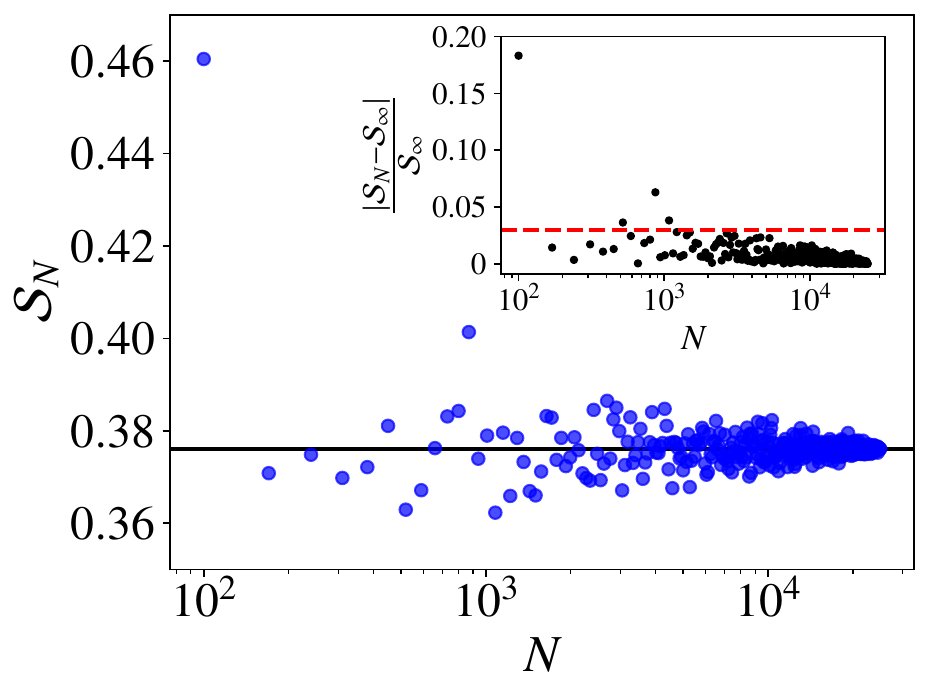}
      \caption{Convergence in the estimate of the topological entropy ($\approx$ 0.376) as we increase the sample size of eigenvalues $N$ at a representative Re $= 10^3$. Inset shows that the relative error in the estimation drops below 3\% as $N \sim \mathcal{O}(10^3)$ which is easily attainable in experiments.}
    \label{fig:err_datapts}
\end{figure}

The topological entropy is determined as the exponential stretching rate of a material curve made up of tracer particles that are passively advected by a smooth flow \cite{newhouse1993estimation}\textemdash an idea that has been utilized to understand the advection dynamics in many open flows \cite{ZIEMNIAK1994123,Hazpra-S,mdkl-nnq1}. Explicitly, the topological entropy $\mathcal{S}$ is characterized by the exponential growth of the length of a material curve that is established asymptotically at $\mathcal{ST} \gg 1$ as (see Fig. \ref{fig:material-line})
\begin{equation}
  l_\mathcal{T} \sim e^{\mathcal{S} \mathcal{T}}.
  \label{eq:S_lag0}
\end{equation}
  The idea is fundamentally rooted in the exponential growth of the number of unstable cycles in a chaotic system that leads to a direct determination of the topological entropy as 
\begin{equation}
  \mathcal{S} = \frac{1}{\mathcal{T}} \ln \bigg(\frac{l_\mathcal{T}}{l_0}\bigg), \quad \mathcal{ST} \gg 1. 
  \label{eq:S_lag}
\end{equation}
To proceed further, we realize that the time evolution of a material line (see Fig.\ref{fig:line_stretching}) manifests as a sum of two independent contributions, namely local deformation and rigid body rotation. This is effected by the velocity gradient tensor, which is written as 
\begin{align*}
  \nabla \bm u = \underbrace{\frac{1}{2}\left(\pdv{u_i}{x_j} + \pdv{u_j}{x_i}\right)}_{\Gamma_{ij}} + \underbrace{\frac{1}{2}\left(\pdv{u_i}{x_j} - \pdv{u_j}{x_i}\right)}_{\Omega_{ij}},
\end{align*}  
where $\bm \Gamma $ is the rate of strain tensor, capturing the symmetric part of the velocity gradient, and $ \bm \Omega$ is the rotational tensor, capturing the anti-symmetric part. This decomposition reflects the general property of any square matrix that it can be written as the sum of a symmetric and an anti-symmetric matrix. This allows us to  evolve the separation vector $\Delta \bm{x}$ between two nearby tracers in time as (see Fig. \ref{fig:line_stretching})
\begin{align}
  \frac{\text{d}}{\text{d}t}\Delta \bm{x} = \Delta \bm{u}, 
  \label{dr/dt}
\end{align}  
with $ \Delta \bm{u} $ as the velocity difference vector between the two tracers that is defined as 
\begin{align*}
  \Delta \bm u = \nabla \bm u \cdot \Delta \bm x =  \bm \Gamma \cdot \Delta \bm x +\frac{1}{2}\bm \omega \times \Delta \bm x,
  \label{eq:delta_u}
\end{align*}
where $\bm{\omega} = \curl{\bm{u}}$ is taken as the local fluid vorticity. As rigid body rotations do not change the length of a vector, the time evolution 
\begin{equation*}
  \frac{\text{d}}{\text{d} t}|\Delta \bm x|^2 = 2 \Delta \bm x \cdot \Delta \bm u =   2  \Delta \bm x \cdot \bm\Gamma \cdot \Delta \bm x, 
  \label{dr2/dt}
\end{equation*}
allows us to rewrite Eq. (\ref{dr/dt}) as 
\begin{equation}
  \frac{\text{d}}{\text{d}t}\Delta \bm{x}  =  \bm\Gamma \cdot \Delta \bm{x},
  \label{eqn:eigval-eq}
\end{equation}
in the eigenbasis of the $\bm \Gamma$ tensor. By discretising the time using a small interval $\tau$, the solution of Eq. \ref{eqn:eigval-eq} at any time index $\nu$ evolves as  
    \begin{eqnarray}
      \Delta z_{\nu+1} &=& \Delta z_\nu\;  e^{\lambda_\nu\tau} \nonumber\\
      \Delta y_{\nu+1} &=& \Delta y_\nu\;  e^{\zeta_\nu\tau} \nonumber\\
      \Delta x_{\nu+1} &=& \Delta x_\nu\;  e^{-(\lambda_\nu + \zeta_\nu)\tau},  
    \end{eqnarray}
    where $\lambda_\nu$ and $\zeta_\nu$ are the top two eigenvalues of $\bm \Gamma$ at the time index $\nu$  with $\lambda_\nu > \zeta_\nu$.  We take  $\text{Tr } \bm \Gamma = 0$ due to the incompressibility of the fluid.  The reader should note that in contrast to our previous work on 2D flows \cite{mdkl-nnq1}, we have to take into account an additional eigenvalue to track the separation vector over time. To proceed further, we 
    take the separation vector resolved as 
    \begin{eqnarray}
      \Delta z_\nu &=& d_\nu\cos\theta_\nu \nonumber\\
      \Delta y_\nu &=& d_\nu\sin\theta_\nu\sin\phi_\nu \nonumber\\
      \Delta x_\nu &=& d_\nu\sin\theta_\nu\cos\phi_\nu, 
    \end{eqnarray}
     where $d_\nu, \theta_\nu$ and $\phi_\nu$ are respectively the separation distance, polar angle and azimuthal angle characterizing the tracer pair at the time index $\nu$. The ratio of pair separations at any  two consecutive times then defines the scaling function $f_\nu$:
    \begin{widetext}
 \begin{equation}
 \frac{d_{\nu+1}}{d_{\nu}} = [ e^{-2(\lambda_\nu + \zeta_\nu)\tau} \sin^2\theta_\nu\cos^2\phi_\nu + e^{2\zeta_\nu\tau}  \sin^2\theta_\nu \sin^2\phi_\nu  + e^{2\lambda_\nu\tau} \cos^2\theta_\nu ]^{1/2} =  f(\lambda_\nu,\zeta_\nu, \theta_\nu, \phi_\nu) \equiv f_\nu.
  \label{eqn:expression_dt} 
\end{equation}
\end{widetext}
  We can now proceed with this idea and track the length of a material line over an arbitrary  time $\mathcal{T} = m \tau$ if the constituting tracer pairs are initially separated by a distance $d_0$  (see Fig. \ref{fig:line_stretching}). Mathematically, this is written as 
\begin{eqnarray}
  l_{\mathcal{T}} = d_0\sum_{i = 1}^{n}\prod_{\nu=0}^{m-1}f_{i\nu} ,
  \label{l_T}
\end{eqnarray}
where the extra index $i$ runs over the tracer pairs. Equation \ref{l_T} together with Eq. \ref{eqn:expression_dt} tells us that the length of the material curve at any time $\mathcal{T} = m\tau$ can be constructed from the distribution of the top two eigenvalues ($\lambda_\nu,\zeta_\nu$) of the strain-rate tensor and the orientation ($\theta_\nu, \phi_\nu$) of the constituent pairs along the material curve. Plugging Eq. \ref{l_T} in Eq. \ref{eq:S_lag} and realizing that $l_0 = n d_0$, we get the topological entropy as
\begin{eqnarray}
  \mathcal{S} &=&  \frac{1}{m\tau}\ln\left[\frac{1}{n}\sum_{i = 1}^{n}\prod_{\nu =0}^{m-1}f_{i\nu}\right] \nonumber\\
  &=& \frac{1}{m\tau}\ln\bigg\langle\prod_{\nu =0}^{m-1} f_{\nu} \bigg\rangle_{\text{pairs}},
  \label{eq:S_lag1}
\end{eqnarray}
where $\langle \cdots \rangle_{\text{pairs}}$ indicates a Lagrangian average over the $n$ tracer pairs constituting the material curve. As shown in ref. \cite{mdkl-nnq1} and explained in Appendix \ref{sec:avg-product-f}, the time-accumulated scaling function $\prod_{\nu=0}^{m-1} f_\nu$  is a log-normal random variable when $m\gg1$. This follows from the positivity of each factor $f_\nu$ and the central limit theorem. The statistical properties then demand that the mean $\langle \prod_{\nu=0}^{m-1} f_\nu \rangle = e^{\mu + \sigma^2/2}$ where $\mu$ and $\sigma^2$ are respectively the mean and variance of the Gaussian distributed $\ln (\prod_{\nu=0}^{m-1} f_\nu)$. This allows us to move the average outside the logarithm in Eq. \ref{eq:S_lag1} to write    
\begin{equation}
  \mathcal{S} = \frac{1}{m\tau} \bigg[ \expval{\ln \prod_{\nu=0}^{m-1} f_{\nu}}_{\text{pairs}} + \frac{\sigma^2}{2}\bigg]. 
 \label{eq:S_lag2} 
\end{equation}
We can further distribute the logarithm of the product as a time average and write as 
\begin{equation}
  \mathcal{S} = \frac{1}{\tau} \bigg[\expval{\ln f}_{\text{pairs,\;time}} + \frac{\sigma^2}{2m}\bigg].  
  \label{eq:S_lag3}
\end{equation}
Equation \ref{eq:S_lag3} is a Lagrangian approach  that requires averaging over both pairs and time which is challenging in experimental turbulence as the number of pairs is prohibitively large. A significant breakthrough is possible if we consider $\tau$ to be the de-correlation time of the eigenvalues provided they exhibit an exponential decay. To that end,   we first define the autocorrelation function 
\begin{equation}
\mathcal{R}(t) = \langle \delta \lambda (\bm x_0, t) \delta \lambda(\bm x_0, t+t_0) \rangle_{\bm x_0, t_0}
\label{eqn:acf}
\end{equation}
where the fluctuation $\delta \lambda (\bm x_0, t) = \lambda(\bm x_0, t) - \langle \lambda(\bm x_0, t) \rangle_{\bm x_0}$. Here $\langle \cdots \rangle_{\bm x_0, t_0}$ denotes an average over a  distribution of spatial locations and initial conditions. In Fig. \ref{fig:decorr_time} we plot $\mathcal{R}(t)$ normalized to its initial value for the top two eigenvalues at a representative $\text{Re} = 10^3$. As $\lambda$ decorrelates slower than $\zeta$, we use its e-folding time to read $\tau \approx 1.5$. We set this as the working protocol to extract $\tau$ that are  plotted in the inset of Fig. \ref{fig:decorr_time}. We are now in a position to simplify Eq. \ref{eq:S_lag3} substantially. Over the time scale $\mathcal{T} \gg \tau$ which signifies the onset of the exponential growth,  the material curve must sample a distribution of independent eigenvalues and angles.
\begin{figure}[ht]
     \includegraphics[width=\linewidth]{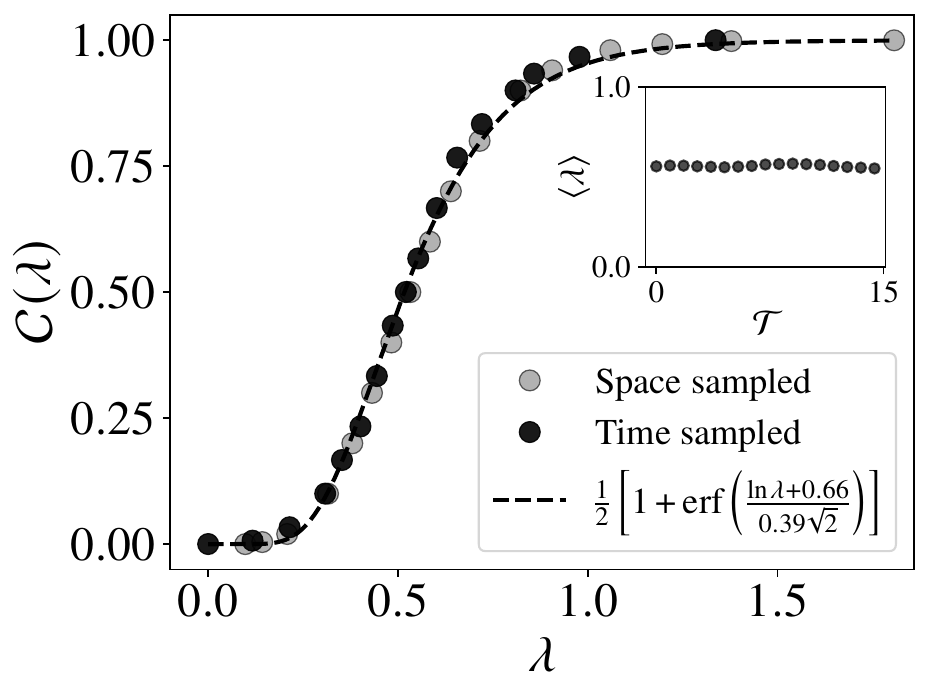}
     \includegraphics[width=\linewidth]{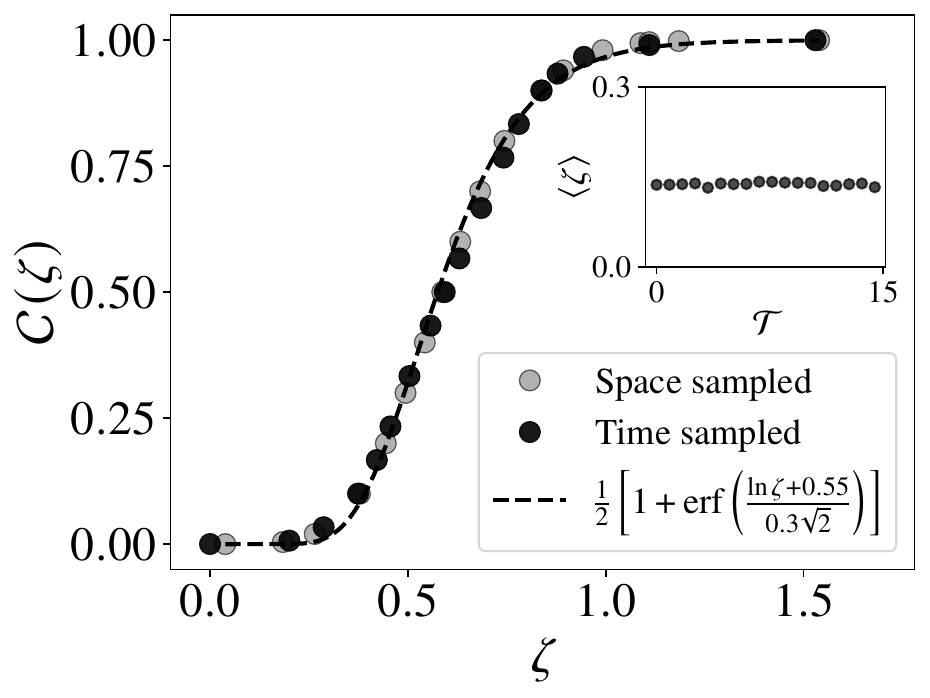}
           \label{fig:lambda_cdf}
    \caption{Cumulative distribution functions of the top two eigenvalues $\lambda$ and $\zeta$ at a representative Re = $10^3$. They are stationary as confirmed by the corresponding insets. By independently sampling the data in both space and time we get an identical distribution that clearly demonstrates ergodicity in our flow. Dashed line in each figure shows a fit with the log-normal distribution. Note: The distribution of $\zeta$ is  shifted on the right by 0.46 as it can sample negative values too. The mean however, is  positive, with $\langle \zeta \rangle \approx 0.14$ at this Re.}
     \label{fig:fig_ergodicity}
\end{figure}\noindent These distributions are stationary which allows us to invoke the ergodic hypothesis \cite{RevModPhys.57.617}  and effectively replace the Lagrangian average in Eq. \ref{eq:S_lag3} with an Eulerian average as
\begin{equation}
  \mathcal{S} =  \frac{1}{\tau} \bigg[\expval{\ln f} + \frac{\sigma^2}{2 m}\bigg].
  \label{eq:S_lag4}
\end{equation}
    
    \noindent where $\langle \ldots \rangle$ denotes averaging over the random variables $\lambda,\zeta,\theta$ and $\phi$. 
For a demonstration of ergodicity in our work, we point the reader to Fig. \ref{fig:fig_ergodicity} where we show that the sampling of eigenvalues in  space and time yield identical distributions. It can be shown with some effort (refer Appendix \ref{sec:sigma-sq-expression}) that the variance 
\begin{equation}
    \sigma^2 = m\bigg[\frac{1}{2} \ln \langle f^2\rangle -  \langle \ln f \rangle\bigg].
\end{equation}
Putting this in Eq. \ref{eq:S_lag4}, we get 
\begin{equation}
  \mathcal{S} = \frac{1}{\tau} \bigg[\frac{1}{2}\expval{\ln f} + \frac{1}{4} \ln \langle f^2 \rangle\bigg].
  \label{eq:S_lag5}
\end{equation}
The partial averages of $\ln f$ and $f^2$ in Eq. \ref{eq:S_lag5} over the angles $\theta$ and $\phi$ can be computed as (refer Appendix \ref{sec:avg-lnf-and-fsq}):  
\begin{widetext}
  \begin{eqnarray}
    \langle \ln f \rangle_{\theta, \phi} &=& -\ln 2 + \frac{2}{\pi}  e^{-(2\lambda+\zeta)\tau} E\bigg(1 - e^{2(\lambda + 2\zeta)\tau} \bigg) - \frac{1}{4} (e^{-2(2\lambda+\zeta)\tau} + e^{-2(\lambda-\zeta)\tau}) + \lambda\tau
     \equiv G(\lambda,\zeta;\tau),\\
    \langle f^2 \rangle_{\theta, \phi} &=& \frac{1}{4} \bigg( e^{-2(\lambda+\zeta)\tau} + e^{2\zeta\tau} + 2 e^{2\lambda\tau} \bigg) \equiv H(\lambda, \zeta;\tau), 
    \label{eqn:H_and_G}
\end{eqnarray}
\end{widetext}
where $E(\cdots)$ is the complete elliptic integral of second kind. Note that the functions $G$ and $H$ depend only on the local eigenvalues $\lambda$ and $\zeta$. Equation \ref{eq:S_lag5} can finally be written as  
\begin{equation}
  \mathcal{S} = \frac{1}{\tau} \bigg[\frac{1}{2} \langle G(\lambda,\zeta;\tau)\rangle  + \frac{1}{4} \ln \langle H(\lambda, \zeta;\tau) \rangle  \bigg].
  \label{eqn:S_eul_final}
\end{equation}
Here $\langle \cdots\rangle$ now denotes averaging over the distributions of the top two eigenvalues $\lambda$ and $\zeta$.
These are readily obtained by diagonalizing the local strain-rate tensor $\mathbf{\Gamma}$.  This approach eliminates the need to track particles moving with the fluid and lends considerable simplification by relying only on the eigenvalue distributions of $\bm \Gamma$. Since we have considered the flow to be ergodic and stationary in nature (see Fig. \ref{fig:fig_ergodicity}), this is enabled by using just a single hot-wire  anemometer \cite{zimmerman2017design} to locally extract the distributions of $\lambda$ and $\zeta$.  In practice, we note that a sample size of $\mathcal{O}(10^3)$ eigenvalues is  sufficient to estimate  $\mathcal{S}$  with a relative error below 3\% (see Fig. \ref{fig:err_datapts}). This greatly brightens the prospects for experimental turbulence where vorticity probes are routinely used \cite{cavo2007performance, wallace2010measurement}.
The same probe can yield a time series data that can be used to construct an auto-correlation whose exponential decay will provide a measure of the de-correlation time $\tau$ (see Eq. \ref{eqn:acf} and Fig. \ref{fig:decorr_time}). In what follows, we provide a direct verification of our result by employing the Lagrangian approach in numerical simulations of 3D turbulence. 

\section{\label{sec:level3} Model and Simulation}
We model the 3D flow using the incompressible Navier-Stokes equations. By expressing convection in terms of fluid vorticity, these equations can be written in the dimensionless form  as \cite{mortensen2016high}:
\begin{eqnarray}
    \pdv{\bm u}{t}  &=& \bm u\times\bm{\omega} - \grad{P} + \frac{1}{\text{Re}} \grad^2\bm u + \bm{F} \nonumber\\
    \div{\bm u} &=& 0, \label{eq:NS_phy}
\end{eqnarray}
where $\bm u$ is the velocity, $\bm{\omega}$ is local vorticity, $P$ is a modified pressure, $\bm{F}$ is a random forcing scheme, and Re is the Reynolds number. To avoid the evaluation of higher-order derivatives, we perform the time evolution of Eq. \ref{eq:NS_phy} in the spectral space. This is done by performing a discrete fourier transform (DFT) of Eq. \ref{eq:NS_phy} and enforcing  incompressibility  to eliminate the pressure $P$.  The model equation in spectral space then reduces to:
\begin{align}
   \pdv{\tilde{\bm u}}{t} = \widetilde{\qty(\bm u \times \bm{\omega})} - \frac{1}{\text{Re}}|\bm k|^2\tilde{\bm u} - \bm k\qty(\frac{\bm k\cdot\widetilde{(\bm u \times \bm{\omega})}}{|\bm k|^2}) + \tilde{\bm{F}}.
\label{eq:NS_Fs}
\end{align}
Here $\widetilde{(\cdots)}$ marks a quantity in $\bm k$-space  with $\bm k \equiv (k_x, k_y, k_z)$ being the 3D wave vector. The non-linear convective term $\widetilde{(\bm{u} \times \bm{\omega})}$  requires implementation of a pseudo-spectral approach where the velocity $\bm u$ and the local vorticity $\bm{\omega}$ are first transferred to the physical space to perform the cross product and then back to the  spectral space for the time integration. Aliasing errors were removed by implementing the standard $2/3$ rule \cite{patterson1971spectral}. In order to reach a steady state, a divergence-free and isotropic random forcing term $\tilde{\bm{F}}_k$ is used to balance the energy dissipation due to fluid viscosity \cite{alvelius1999random}. In practice, this forcing is restricted to a narrow band of wave numbers with $|\bm{k}|\in [3,7]$. As the velocities at smaller scales evolve unforced, the corresponding Lagrangian acceleration, particularly at the Kolmogorov dissipative scale, are smoothed out. This scale separation which gets progressively better at large Reynolds number lends a marginal role to the nature of forcing in the Lagrangian evolution \cite{eswaran1988, chen2015lagrangian, yeung1989lagrangian}. A parallel code was written in C programming language using message passing interface and the DFTs were performed using the FFTW Library \cite{mortensen2016high}. We employed a $2^{\text{nd}}$ order Crank-Nicolson scheme to integrate Eq. \ref{eq:NS_Fs} with a time step  $\Delta t = 0.005$. This way, we were able to maintain the Courant-Friedrichs-Lewy criterion for numerical stability over the entire range of Reynolds numbers reported in our work. The numerical simulations were performed on a triply periodic cubic grid having $512^3$ grid points with a side length of $2\pi$. The details of our Lagrangian measurement are described next.

\begin{figure}[h]
    \raggedright
    \includegraphics[width=\columnwidth]{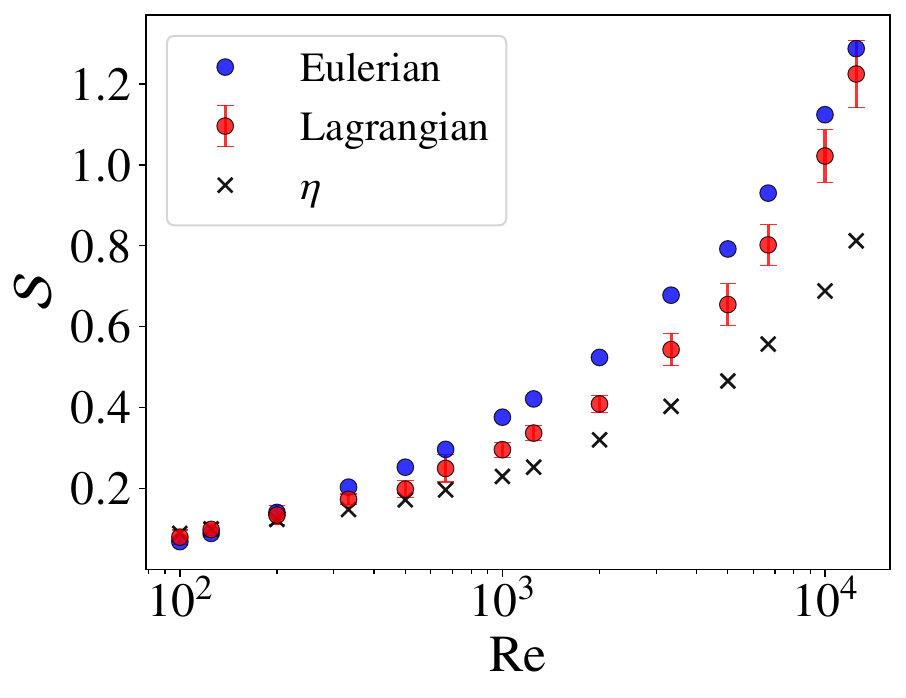}
    \caption{Topological entropy vs. Reynolds number. The Lagrangian and Eulerian measures show reasonable agreement over two decades of variation in Re. For a comparison, we add a plot of the Lyapunov exponent $\eta$ that is described in Appendix \ref{sec:metric-s}. As expected, $\eta \leq \mathcal{S}$ at all Re \cite{ott2002chaos}.}
    \label{fig:top_ent}
\end{figure}

In the steady state of turbulence, we introduce a material line composed of $7\times10^6$ tracer particles that are passively advected using the following equation:
\begin{align}
    \frac{\text{d}}{\text{d}t}\bm{r}_i(t) = \bm{u}(\bm{r}_i(t)).
\end{align}
Here $\bm{r}_i(t)$ and $\bm{u}(\bm{r}_i(t))$ respectively denote the location and velocity of the $i^{\text{th}}$ passive tracer. The tracer velocity is computed by projecting the fluid velocity at the tracer location through a linear interpolation algorithm. The density of tracers in the initial configuration is  large enough to keep the tracer pairs within the Kolmogorov dissipative scale at all times. At any time, the length of the material curve is estimated by summing the distances between the consecutive tracers. The exponential growth rate is then used to compute the Lagrangian measure  that is presented in Eq.\,\ref{eq:S_lag}. As a working protocol, we always wait until $\mathcal{ST} > 10$ before reading the exponential growth of the curve. This order of magnitude separation between the timescales $\mathcal{T}$ and $1/\mathcal{S}$ allows the material curve to sample the entire box as shown in Fig. \ref{fig:material-line}. For each Re, we generate 10 independent realizations of stretching by varying the initial distribution of applied forcing $\bm F$ and setting the initial configurations of the material curve as spiral, straight line and semicircle.  The resulting mean growth rate along with the error bars are taken as the Lagrangian measure shown in Fig. \ref{fig:top_ent}. The particle tracking described here presents a significant technical difficulty in experimental turbulence as their trajectories are poorly resolved due to chaos. In the following, we compute the topological entropy under the Eulerian framework developed in this paper. 

On achieving the steady state of the flow, we obtain the distributions of eigenvalues $\lambda$ and $\zeta$  by diagonalizing the local strain-rate tensor $\bm \Gamma$. These distributions enable the calculation of topological entropy by an Eulerian measure provided in Eq.\,\ref{eqn:S_eul_final}.  
The result is plotted in Fig. \ref{fig:top_ent} to compare with the Lagrangian measure that was computed earlier. The two measures are in reasonable agreement with the relative error remaining below 20\% over two decades of Re. For a statistical theory that is devoid of free parameters this constitutes  a clear demonstration of the working principle that is independent of the  underlying model or experimental data.  To an experimentalist, this holds great promise as the topological entropy can be computed entirely from the local velocity gradient tensor —a quantity that is easily obtained from traditional probes. We also add the graph of a related quantity, namely the Lyapunov exponent $\eta$ as a third plot in Fig. \ref{fig:top_ent}. In Appendix \ref{sec:metric-s} we show how this can be calculated over a distribution of tracer pairs. As expected, $\eta$ is bounded above by $\mathcal{S}$ \cite{ott2002chaos}. Notice that  $\mathcal{S}$ increases non-linearly with increasing Re. Intuitively, one expects this from the positive correlation of enstrophy density $|\bm \omega |^2$ with Re that may intensify local velocity gradients that aid in mixing of passive tracers. In the following, we conclude the manuscript by highlighting the experimental utility and natural extensions of this work.

\section{\label{sec:level4}Conclusion}
In this work we provide a theory to calculate the topological entropy under the Eulerian framework in a 3D setting. The result is in reasonable agreement with the Lagrangian measure directly obtained from numerical simulations of inertial turbulence, over two decades of Re. We eliminate the need for Lagrangian tracking of tracers advected by chaotic flows and provide a simple Eulerian formulation that is easily amenable in experiments. Our theory does not require free parameters to be fitted against experimental data and therefore constitute a demonstration of the  principle governing the deformation of a  fluid material line.  This lends considerable utility  to experimental turbulence as the topological entropy can be computed directly from the local properties of the flow. The result is a global measure of complexity unlike the Lyapunov exponent which requires the trajectories to remain much closer than the linear scale of the flow \textemdash a condition impractical to maintain in turbulence. Within the domain of industrial flows, our results may apply to pharmaceutical manufacturing \cite{crouter2019methods, ebrahimi2019application}, fuel mixing in combustion chambers \cite{zoby2011turbulent} and even coolant mixing in nuclear reactors \cite{li2021cfd}. Extending to geophysical systems, our results may apply to oceanic and atmospheric flows when vertical shear is high and localized sub-mesoscale processes are not significant. In such cases, our 3D calculations can be used for accurately determining the topological entropy. This is particularly relevant in mesoscale turbulence that is often  characterized by length scales of $\mathcal{O}(10 - 100 \text{ km})$  and timescales of $\mathcal{O}(\text{weeks} )$ over which the assumptions of homogeneity and ergodicity generally hold \cite{unpub-geophys-comm}. The strain-rate tensor will then dominate the dynamics of the flow and our theory should reliably capture the stretching of material lines \cite{srinivasan2023forward}. 
However, caution must be exercised to avoid factors such as seasonal and topographic variations, island chains, internal waves and strong gradient fronts in oceanic flows that may render these assumptions invalid \cite{d2018ocean, jagannathan2021boundary, srinivasan2023forward}. We believe our results reported here will significantly add to the current understanding of transport and mixing of passive tracers in 3D turbulent flows. 
\section*{Acknowledgements}
We thank Arjun Jagannathan, Department of Ocean Engineering at IIT Madras for comments and discussions.

\appendix

\section{\label{sec:avg-product-f} The distribution of $\prod_{\nu=0}^{m-1} f_\nu$}
\noindent The factors $f_\nu > 0$ are independent and identically distributed. The logarithm of the product can therefore be written as a real random variable:
\begin{equation}
  \ln (\prod_{\nu=0}^{m-1} f_\nu) = \sum_{\nu=0}^{m-1} \ln f_\nu = x.
\end{equation}
For $m \gg 1$,  central limit theorem guarantees that $x$ must be normally distributed with mean $\mu$ and variance $\sigma^2$. This makes the product $\prod_{\nu=0}^{m-1} f_\nu$ lognormal, with the average:  
\begin{equation}
\bigg \langle \prod_{\nu=0}^{m-1} f_\nu \bigg\rangle = \langle e^x \rangle = \int_{-\infty}^{\infty} \frac{e^{x-(x-\mu)^2/2\sigma^2}}{\sqrt{2\pi \sigma^2}} \;\text{d}x = e^{\mu + \sigma^2/2}.
\label{eq:1st-moment-prod-f}
\end{equation}

\section{\label{sec:sigma-sq-expression} Writing $\sigma^2$ using moments of $f$ and $\ln f$}
\noindent The mean is already worked out in Eq. \ref{eq:1st-moment-prod-f}. We work out the second moment as 
\begin{equation}
\bigg \langle \bigg(\prod_{\nu=0}^{m-1} f_\nu \bigg)^2\bigg\rangle = \langle e^{2x} \rangle = \int_{-\infty}^{\infty} \frac{e^{2x-(x-\mu)^2/2\sigma^2}}{\sqrt{2\pi \sigma^2}} \;\text{d}x = e^{2(\mu + \sigma^2)}.
\label{eq:2nd-moment-prod-f}
\end{equation}
Combining \ref{eq:1st-moment-prod-f} and \ref{eq:2nd-moment-prod-f}, the variance is easily realized as  
\begin{equation}
  \text{Var}\bigg(\prod_{\nu=0}^{m-1} f_\nu \bigg) = e^{2\mu + \sigma^2}(e^{\sigma^2} - 1) = \langle f^2 \rangle^m - \langle f \rangle^{2m}. 
  \label{eq:var-prod-f}
\end{equation}
The last equality in \ref{eq:var-prod-f} arises from the fact that the factors $f_\nu$ are independent and identically distributed random variables due to the homogeneity and stationarity of the flow. This implies that the associated mean and variance are simply scaled by the number of factors in the product: 
\begin{eqnarray}
    \mu &=& m \langle \ln f\rangle, \nonumber\\
    \sigma^2 &=& m \text{Var}(\ln f).
\end{eqnarray}
Clearly for a large waiting time or $m\gg1$, we can take $e^{\sigma^2} \gg 1$. Under this limit, the logarithm of Eq. \ref{eq:var-prod-f} gives
\begin{eqnarray}
  \sigma^2 &=& \frac{1}{2} \ln [\langle f^2 \rangle^m - \langle f \rangle^{2m}] - m \langle \ln f \rangle.
  \label{<+label+>}
\end{eqnarray}
Due to the positivity of variance, we realize that the ratio $\langle f \rangle^{2m} / \langle f^2 \rangle^m \ll 1$ when $m\gg1$. This gives us the limit
\begin{equation}
    \sigma^2 = \frac{m}{2}\ln \langle f^2 \rangle - m \langle \ln f \rangle.
\end{equation}
We plug this back in the manuscript Eq. \ref{eq:S_lag4} to proceed further.\\

\section{\label{sec:avg-lnf-and-fsq}Computation of   $\expval{\ln f}_{\theta, \phi}$ and $\expval{f^2}_{\theta, \phi}$}
To compute these, we consider the angles $\theta$ and $\phi$ to have uniform random distributions in the range $[0, \pi/2]$ which is also supported by our numerical data. By squaring $f$ defined in Eq. \ref{eqn:expression_dt} and 
realizing that the averages $\expval{\sin^2 x} = \expval{\cos^2 x} = 1/2$ for $0\leq x\leq \pi/2$, we can immediately write down the angle averaged $f^2$ as:
\begin{eqnarray}
    \expval{f^2}_{\theta, \phi} &=& \frac{1}{4} \bigg( e^{-2(\lambda+\zeta)\tau} + e^{2\zeta\tau} + 2 e^{2\lambda\tau} \bigg).
    \label{eqn:avf-fsq}
\end{eqnarray}
Equation \ref{eqn:avf-fsq} is ready to be substituted in Eq. \ref{eq:S_lag5}. 
The logarithm of $f$ is angle averaged in two steps. First, we average over $\theta$: 
\begin{widetext}
    \begin{equation}
    \langle \ln f \rangle_{\theta} = \expval{\ln\cos\theta}_{\theta}  + \frac{1}{2}\expval{\ln (1+\qty[e^{-2(2\lambda + \zeta)\tau} \cos^2\phi + e^{2(\zeta - \lambda)\tau} \sin^2\phi]\tan^2\theta)}_\theta + \lambda \tau .
    \label{eqn:thetaAvg_lnf}
\end{equation}
\end{widetext}
The first term in the RHS of Eq. \ref{eqn:thetaAvg_lnf} is merely 
\begin{equation}
    \expval{\ln \cos \theta}_{\theta} = \frac{2}{\pi}\int_0^{\pi / 2} \ln \cos \theta \;   \text{d}\theta = - \text{ln} \;2 .
    \label{eqn:ln-cos}
\end{equation}
With the substitutions $\tan \theta = x$, $e^{-2(2\lambda + \zeta)\tau} = \alpha$, $e^{2(\zeta - \lambda)\tau} = \beta$ and $\alpha\cos^2\phi + \beta \sin^2\phi = C$, the second term in the RHS of Eq. \ref{eqn:thetaAvg_lnf} can be written as an integral
\begin{equation}
    \frac{1}{2}\langle \ln(1 + C \tan^2 \theta)\rangle_{\theta} = \frac{1}{\pi}\int_0^{\infty} \frac{\ln (1+ C x^2)}{1+x^2} \text dx = I.
      \label{eqn:thetaAvg1}
\end{equation}
The integral $I$ in Eq. \ref{eqn:thetaAvg1} is a Serret-integral \cite{GradshteynRyzhik2000} that can be solved through the technique of contour integration. To do this, we solve the following complex integral on the contour shown in Fig.\ref{fig:contour} using Cauchy's residue theorem: 
\begin{figure}[h!]
  \centering
    \includegraphics[width=0.635\columnwidth]{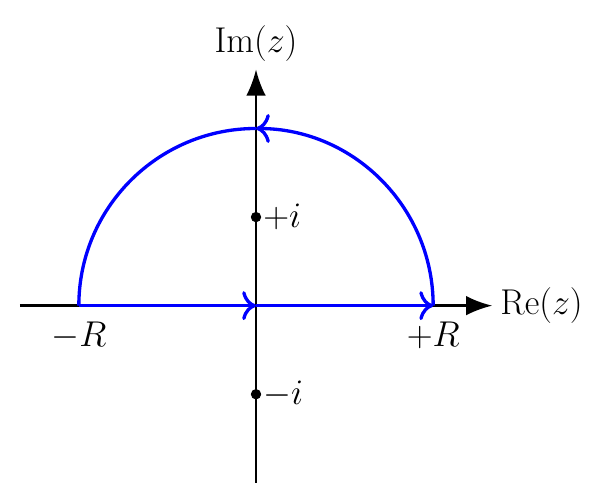}
  \caption{The blue semicircle of radius $R$ represents a closed-contour including the pole $+i$ along which the integral $J$ has been computed.}
  \label{fig:contour}
\end{figure}
\noindent
\begin{equation}
J = \frac{1}{\pi}\oint_C \frac{\ln(i+z\sqrt{C})}{1+z^2} \text dz = \frac{i\pi}{2} + \ln(1+\sqrt{C}).\label{eqn:J1}
   \end{equation}
Along the contour in Fig. \ref{fig:contour}, the complex integral is
\begin{align}
\nonumber
J &= \frac{1}{\pi}\bigg[\int_{-R}^R \frac{\ln(i+x\sqrt{C})}{1+x^2} \text dx\\
&\quad \quad \quad \quad \quad + \int_{\theta=0}^\pi \frac{\ln\left(i+R e^{i\theta}\sqrt{C}\right)}{1+R^2 e^{2i\theta}} R i e^{i\theta}\text d\theta\bigg]. \label{eqn:J2}
\end{align}
In the limit $R \rightarrow \infty$, the magnitude of the integrand
\begin{align*}
\left|\frac{\ln\left(i+R e^{i\theta} \sqrt{C}\right) R i e^{i\theta}}{1+R^2 e^{2i\theta}}\right| &\longrightarrow 0.
\end{align*}
Thus the arc integral vanishes and Eq. \ref{eqn:J2} reduces to 
\begin{eqnarray}
J &=& \frac{1}{\pi}\qty[ \int_{-\infty}^0 \frac{\ln(i+x\sqrt{C})}{1+x^2}\text dx + \int_0^{\infty} \frac{\ln(i+x\sqrt{C})}{1+x^2}\text dx] \nonumber\\
&=& \frac{1}{\pi}\int_0^{\infty} \frac{\ln(-1) + \ln(1 + x^2C)}{1+x^2}\text dx\nonumber\\
&=& \frac{i\pi}{2} + I. 
\label{eqn:J3}
\end{eqnarray}
Comparing Eqs. \ref{eqn:J1} and \ref{eqn:J3}, the integral in Eq. \ref{eqn:thetaAvg1} becomes
\begin{equation}
  I =  \ln (1+\sqrt{C}).
  \label{eqn:thetaAvg_final}
\end{equation}
Plugging this in Eq. \ref{eqn:thetaAvg_lnf} we get the $\theta$-averaged $f$:
\begin{align}
    \langle \ln f \rangle_{\theta} &= -\ln 2  + \ln (1+\sqrt{C})  + \lambda \tau . \label{eqn:thetaAvg_lnf_final}
\end{align}
The only term on the RHS of Eq. \ref{eqn:thetaAvg_lnf_final} that depends on $\phi$ is $\ln(1+\sqrt{C})$. Therefore, to complete the $\phi$-average of $\langle \ln f \rangle_{\theta}$, we just need  the average 
\begin{widetext}
\begin{eqnarray}
\bigg\langle \ln (1 + \sqrt{C}) \bigg \rangle_{\phi} &=& \frac{2}{\pi} \int_{0}^{\pi/2}\ln\qty(1+\sqrt{\alpha \cos^2\phi + \beta \sin^2\phi})\text d\phi\nonumber\\
&=& \frac{2}{\pi} \int_{0}^{\pi/2} \bigg[\sqrt{\alpha \sin^2\phi + \beta\cos^2\phi}-\frac{1}{2} (\alpha \cos^2\phi + \beta \sin^2\phi) + \mathcal{O}\bigg((\alpha \cos^2\phi + \beta \sin^2\phi)^{3/2}\bigg)\bigg]\text{d}\phi \nonumber\\
&\approx& \frac{2}{\pi} \bigg[ \sqrt{\alpha} E\bigg(1 - \frac{\beta}{\alpha} \bigg) - \frac{\pi}{8} (\alpha+\beta)\bigg].
\label{eqn:phi-avg-ln-1-plus-rootc}
\end{eqnarray}
\end{widetext}
Notice that we approximated the above integral by expanding  the logarithm up to the quadratic term as both $\alpha$ and $\beta$ are less than unity ($\lambda$ and $\zeta$ are the top two eigenvalues of the $\bm \Gamma$ tensor). The function $E(\cdots)$ is the complete elliptic integral of the $2^{\text{nd}}$ kind. Putting back the values of $\alpha = e^{-2(2\lambda + \zeta)\tau}$ and $\beta = e^{2(\zeta - \lambda)\tau}$ in Eq.\ref{eqn:phi-avg-ln-1-plus-rootc}, we can finally write the $\phi$-average of Eq. \ref{eqn:thetaAvg_lnf_final} as 
\begin{widetext}
    \begin{equation}
        \langle \ln f \rangle_{\theta,\phi} = -\ln 2 + \frac{2}{\pi}  e^{-(2\lambda+\zeta)\tau} E\bigg(1 - e^{2(\lambda + 2\zeta)\tau} \bigg) - \frac{1}{4} (e^{-2(2\lambda+\zeta)\tau} + e^{-2(\lambda-\zeta)\tau}) + \lambda\tau .
    \end{equation}
\end{widetext}

\section{\label{sec:metric-s} Lyapunov exponent $\eta$}
To compute this quantity, we first obtain the finite time  Lyapunov exponent from the exponential divergence of a tracer pair that starts out infinitesimally close with separation $d_0$:
\begin{equation}
\Lambda_{\text{FT}} (t)  = \frac{1}{t} \ln \bigg(\frac{d_t} {d_0} \bigg) 
\end{equation}
Averaging this over long time gives us the Lyapunov exponent of this pair
\begin{equation}
\Lambda = \lim_{T \to \infty}\frac{1}{T} \int_0^\infty \Lambda_{\text{FT}} (t) \text{d}t
\end{equation}
Finally, we average the positive values of this exponent over the distribution of $n$ tracer pairs to get the Lyapunov exponent
\begin{equation}
\eta = \frac{1}{n} \sum_{i=1}^n \Lambda_i,\quad \Lambda_i \geq 0 
\end{equation}
This is plotted in Fig. \ref{fig:top_ent} of the manuscript. Clearly, $\eta$ is bounded  above by the topological entropy at all Re \cite{ott2002chaos}.
\bibliography{manuscript}

\begin{thebibliography}{52}%
\makeatletter
\providecommand \@ifxundefined [1]{%
 \@ifx{#1\undefined}
}%
\providecommand \@ifnum [1]{%
 \ifnum #1\expandafter \@firstoftwo
 \else \expandafter \@secondoftwo
 \fi
}%
\providecommand \@ifx [1]{%
 \ifx #1\expandafter \@firstoftwo
 \else \expandafter \@secondoftwo
 \fi
}%
\providecommand \natexlab [1]{#1}%
\providecommand \enquote  [1]{``#1''}%
\providecommand \bibnamefont  [1]{#1}%
\providecommand \bibfnamefont [1]{#1}%
\providecommand \citenamefont [1]{#1}%
\providecommand \href@noop [0]{\@secondoftwo}%
\providecommand \href [0]{\begingroup \@sanitize@url \@href}%
\providecommand \@href[1]{\@@startlink{#1}\@@href}%
\providecommand \@@href[1]{\endgroup#1\@@endlink}%
\providecommand \@sanitize@url [0]{\catcode `\\12\catcode `\$12\catcode
  `\&12\catcode `\#12\catcode `\^12\catcode `\_12\catcode `\%12\relax}%
\providecommand \@@startlink[1]{}%
\providecommand \@@endlink[0]{}%
\providecommand \url  [0]{\begingroup\@sanitize@url \@url }%
\providecommand \@url [1]{\endgroup\@href {#1}{\urlprefix }}%
\providecommand \urlprefix  [0]{URL }%
\providecommand \Eprint [0]{\href }%
\providecommand \doibase [0]{https://doi.org/}%
\providecommand \selectlanguage [0]{\@gobble}%
\providecommand \bibinfo  [0]{\@secondoftwo}%
\providecommand \bibfield  [0]{\@secondoftwo}%
\providecommand \translation [1]{[#1]}%
\providecommand \BibitemOpen [0]{}%
\providecommand \bibitemStop [0]{}%
\providecommand \bibitemNoStop [0]{.\EOS\space}%
\providecommand \EOS [0]{\spacefactor3000\relax}%
\providecommand \BibitemShut  [1]{\csname bibitem#1\endcsname}%
\let\auto@bib@innerbib\@empty
\bibitem [{\citenamefont {Zhou}(2021)}]{zhou2021turbulence}%
  \BibitemOpen
  \bibfield  {author} {\bibinfo {author} {\bibfnamefont {Y.}~\bibnamefont
  {Zhou}},\ }\bibfield  {title} {\bibinfo {title} {Turbulence theories and
  statistical closure approaches},\ }\href@noop {} {\bibfield  {journal}
  {\bibinfo  {journal} {Physics Reports}\ }\textbf {\bibinfo {volume} {935}},\
  \bibinfo {pages} {1} (\bibinfo {year} {2021})}\BibitemShut {NoStop}%
\bibitem [{\citenamefont {Wensink}\ \emph {et~al.}(2012)\citenamefont
  {Wensink}, \citenamefont {Dunkel}, \citenamefont {Heidenreich}, \citenamefont
  {Drescher}, \citenamefont {Goldstein}, \citenamefont {L{\"o}wen},\ and\
  \citenamefont {Yeomans}}]{wensink2012meso}%
  \BibitemOpen
  \bibfield  {author} {\bibinfo {author} {\bibfnamefont {H.~H.}\ \bibnamefont
  {Wensink}}, \bibinfo {author} {\bibfnamefont {J.}~\bibnamefont {Dunkel}},
  \bibinfo {author} {\bibfnamefont {S.}~\bibnamefont {Heidenreich}}, \bibinfo
  {author} {\bibfnamefont {K.}~\bibnamefont {Drescher}}, \bibinfo {author}
  {\bibfnamefont {R.~E.}\ \bibnamefont {Goldstein}}, \bibinfo {author}
  {\bibfnamefont {H.}~\bibnamefont {L{\"o}wen}},\ and\ \bibinfo {author}
  {\bibfnamefont {J.~M.}\ \bibnamefont {Yeomans}},\ }\bibfield  {title}
  {\bibinfo {title} {Meso-scale turbulence in living fluids},\ }\href@noop {}
  {\bibfield  {journal} {\bibinfo  {journal} {Proceedings of the National
  Academy of Sciences}\ }\textbf {\bibinfo {volume} {109}},\ \bibinfo {pages}
  {14308} (\bibinfo {year} {2012})}\BibitemShut {NoStop}%
\bibitem [{\citenamefont {Canuto}(2009)}]{canuto2009turbulence}%
  \BibitemOpen
  \bibfield  {author} {\bibinfo {author} {\bibfnamefont {V.}~\bibnamefont
  {Canuto}},\ }\bibfield  {title} {\bibinfo {title} {Turbulence in
  astrophysical and geophysical flows},\ }\href@noop {} {\bibfield  {journal}
  {\bibinfo  {journal} {Interdisciplinary aspects of turbulence}\ ,\ \bibinfo
  {pages} {107}} (\bibinfo {year} {2009})}\BibitemShut {NoStop}%
\bibitem [{\citenamefont {Canuto}\ and\ \citenamefont
  {Christensen-Dalsgaard}(1998)}]{canuto1998turbulence}%
  \BibitemOpen
  \bibfield  {author} {\bibinfo {author} {\bibfnamefont {V.}~\bibnamefont
  {Canuto}}\ and\ \bibinfo {author} {\bibfnamefont {J.}~\bibnamefont
  {Christensen-Dalsgaard}},\ }\bibfield  {title} {\bibinfo {title} {Turbulence
  in astrophysics: Stars},\ }\href@noop {} {\bibfield  {journal} {\bibinfo
  {journal} {Annual review of fluid mechanics}\ }\textbf {\bibinfo {volume}
  {30}},\ \bibinfo {pages} {167} (\bibinfo {year} {1998})}\BibitemShut
  {NoStop}%
\bibitem [{\citenamefont {Reynolds}(1883)}]{reynolds1883xxix}%
  \BibitemOpen
  \bibfield  {author} {\bibinfo {author} {\bibfnamefont {O.}~\bibnamefont
  {Reynolds}},\ }\bibfield  {title} {\bibinfo {title} {Xxix. an experimental
  investigation of the circumstances which determine whether the motion of
  water shall be direct or sinuous, and of the law of resistance in parallel
  channels},\ }\href@noop {} {\bibfield  {journal} {\bibinfo  {journal}
  {Philosophical Transactions of the Royal society of London}\ ,\ \bibinfo
  {pages} {935}} (\bibinfo {year} {1883})}\BibitemShut {NoStop}%
\bibitem [{\citenamefont {Kolmogorov}(1968)}]{kolmogorov1968local}%
  \BibitemOpen
  \bibfield  {author} {\bibinfo {author} {\bibfnamefont {A.}~\bibnamefont
  {Kolmogorov}},\ }\bibfield  {title} {\bibinfo {title} {Local structure of
  turbulence in an incompressible viscous fluid at very high reynolds
  numbers},\ }\href@noop {} {\bibfield  {journal} {\bibinfo  {journal} {Soviet
  Physics Uspekhi}\ }\textbf {\bibinfo {volume} {10}},\ \bibinfo {pages} {734}
  (\bibinfo {year} {1968})}\BibitemShut {NoStop}%
\bibitem [{\citenamefont {Kolmogorov}(1991)}]{kolmogorov1991dissipation}%
  \BibitemOpen
  \bibfield  {author} {\bibinfo {author} {\bibfnamefont {A.~N.}\ \bibnamefont
  {Kolmogorov}},\ }\bibfield  {title} {\bibinfo {title} {Dissipation of energy
  in the locally isotropic turbulence},\ }\href@noop {} {\bibfield  {journal}
  {\bibinfo  {journal} {Proceedings of the Royal Society of London. Series A:
  Mathematical and Physical Sciences}\ }\textbf {\bibinfo {volume} {434}},\
  \bibinfo {pages} {15} (\bibinfo {year} {1991})}\BibitemShut {NoStop}%
\bibitem [{\citenamefont {Fefferman}(2006)}]{fefferman2006existence}%
  \BibitemOpen
  \bibfield  {author} {\bibinfo {author} {\bibfnamefont {C.~L.}\ \bibnamefont
  {Fefferman}},\ }\bibfield  {title} {\bibinfo {title} {Existence and
  smoothness of the navier-stokes equation},\ }\href@noop {} {\bibfield
  {journal} {\bibinfo  {journal} {The millennium prize problems}\ }\textbf
  {\bibinfo {volume} {57}},\ \bibinfo {pages} {22} (\bibinfo {year}
  {2006})}\BibitemShut {NoStop}%
\bibitem [{\citenamefont {Alexakis}\ and\ \citenamefont
  {Biferale}(2018)}]{alexakis2018cascades}%
  \BibitemOpen
  \bibfield  {author} {\bibinfo {author} {\bibfnamefont {A.}~\bibnamefont
  {Alexakis}}\ and\ \bibinfo {author} {\bibfnamefont {L.}~\bibnamefont
  {Biferale}},\ }\bibfield  {title} {\bibinfo {title} {Cascades and transitions
  in turbulent flows},\ }\href@noop {} {\bibfield  {journal} {\bibinfo
  {journal} {Physics Reports}\ }\textbf {\bibinfo {volume} {767}},\ \bibinfo
  {pages} {1} (\bibinfo {year} {2018})}\BibitemShut {NoStop}%
\bibitem [{\citenamefont {Falkovich}\ \emph {et~al.}(2001)\citenamefont
  {Falkovich}, \citenamefont {Gawedzki},\ and\ \citenamefont
  {Vergassola}}]{falkovich2001particles}%
  \BibitemOpen
  \bibfield  {author} {\bibinfo {author} {\bibfnamefont {G.}~\bibnamefont
  {Falkovich}}, \bibinfo {author} {\bibfnamefont {K.}~\bibnamefont
  {Gawedzki}},\ and\ \bibinfo {author} {\bibfnamefont {M.}~\bibnamefont
  {Vergassola}},\ }\bibfield  {title} {\bibinfo {title} {Particles and fields
  in fluid turbulence},\ }\href@noop {} {\bibfield  {journal} {\bibinfo
  {journal} {Reviews of modern Physics}\ }\textbf {\bibinfo {volume} {73}},\
  \bibinfo {pages} {913} (\bibinfo {year} {2001})}\BibitemShut {NoStop}%
\bibitem [{\citenamefont {Wolf}\ \emph {et~al.}(1985)\citenamefont {Wolf},
  \citenamefont {Swift}, \citenamefont {Swinney},\ and\ \citenamefont
  {Vastano}}]{wolf1985determining}%
  \BibitemOpen
  \bibfield  {author} {\bibinfo {author} {\bibfnamefont {A.}~\bibnamefont
  {Wolf}}, \bibinfo {author} {\bibfnamefont {J.~B.}\ \bibnamefont {Swift}},
  \bibinfo {author} {\bibfnamefont {H.~L.}\ \bibnamefont {Swinney}},\ and\
  \bibinfo {author} {\bibfnamefont {J.~A.}\ \bibnamefont {Vastano}},\
  }\bibfield  {title} {\bibinfo {title} {Determining lyapunov exponents from a
  time series},\ }\href@noop {} {\bibfield  {journal} {\bibinfo  {journal}
  {Physica D: nonlinear phenomena}\ }\textbf {\bibinfo {volume} {16}},\
  \bibinfo {pages} {285} (\bibinfo {year} {1985})}\BibitemShut {NoStop}%
\bibitem [{\citenamefont {Oseledec}(1968)}]{oseledec1968multiplicative}%
  \BibitemOpen
  \bibfield  {author} {\bibinfo {author} {\bibfnamefont {V.~I.}\ \bibnamefont
  {Oseledec}},\ }\bibfield  {title} {\bibinfo {title} {A multiplicative ergodic
  theorem, lyapunov characteristic numbers for dynamical systems},\ }\href@noop
  {} {\bibfield  {journal} {\bibinfo  {journal} {Transactions of the Moscow
  Mathematical Society}\ }\textbf {\bibinfo {volume} {19}},\ \bibinfo {pages}
  {197} (\bibinfo {year} {1968})}\BibitemShut {NoStop}%
\bibitem [{\citenamefont {Wolf}\ \emph {et~al.}(1986)\citenamefont {Wolf} \emph
  {et~al.}}]{wolf1986quantifying}%
  \BibitemOpen
  \bibfield  {author} {\bibinfo {author} {\bibfnamefont {A.}~\bibnamefont
  {Wolf}} \emph {et~al.},\ }\bibfield  {title} {\bibinfo {title} {Quantifying
  chaos with lyapunov exponents},\ }\href@noop {} {\bibfield  {journal}
  {\bibinfo  {journal} {Chaos}\ }\textbf {\bibinfo {volume} {16}},\ \bibinfo
  {pages} {285} (\bibinfo {year} {1986})}\BibitemShut {NoStop}%
\bibitem [{\citenamefont {Meneveau}\ and\ \citenamefont
  {Sreenivasan}(1987)}]{meneveau1987multifractal}%
  \BibitemOpen
  \bibfield  {author} {\bibinfo {author} {\bibfnamefont {C.}~\bibnamefont
  {Meneveau}}\ and\ \bibinfo {author} {\bibfnamefont {K.~R.}\ \bibnamefont
  {Sreenivasan}},\ }\bibfield  {title} {\bibinfo {title} {The multifractal
  spectrum of the dissipation field in turbulent flows},\ }\href@noop {}
  {\bibfield  {journal} {\bibinfo  {journal} {Nuclear Physics B-Proceedings
  Supplements}\ }\textbf {\bibinfo {volume} {2}},\ \bibinfo {pages} {49}
  (\bibinfo {year} {1987})}\BibitemShut {NoStop}%
\bibitem [{\citenamefont {Adler}\ \emph {et~al.}(1965)\citenamefont {Adler},
  \citenamefont {Konheim},\ and\ \citenamefont
  {McAndrew}}]{adler1965topological}%
  \BibitemOpen
  \bibfield  {author} {\bibinfo {author} {\bibfnamefont {R.~L.}\ \bibnamefont
  {Adler}}, \bibinfo {author} {\bibfnamefont {A.~G.}\ \bibnamefont {Konheim}},\
  and\ \bibinfo {author} {\bibfnamefont {M.~H.}\ \bibnamefont {McAndrew}},\
  }\bibfield  {title} {\bibinfo {title} {Topological entropy},\ }\href@noop {}
  {\bibfield  {journal} {\bibinfo  {journal} {Transactions of the American
  Mathematical Society}\ }\textbf {\bibinfo {volume} {114}},\ \bibinfo {pages}
  {309} (\bibinfo {year} {1965})}\BibitemShut {NoStop}%
\bibitem [{\citenamefont {Ott}(2002)}]{ott2002chaos}%
  \BibitemOpen
  \bibfield  {author} {\bibinfo {author} {\bibfnamefont {E.}~\bibnamefont
  {Ott}},\ }\href@noop {} {\emph {\bibinfo {title} {Chaos in Dynamical
  Systems}}},\ \bibinfo {edition} {2nd}\ ed.\ (\bibinfo  {publisher} {Cambridge
  University Press},\ \bibinfo {year} {2002})\BibitemShut {NoStop}%
\bibitem [{\citenamefont {Newhouse}\ and\ \citenamefont
  {Pignataro}(1993)}]{newhouse1993estimation}%
  \BibitemOpen
  \bibfield  {author} {\bibinfo {author} {\bibfnamefont {S.}~\bibnamefont
  {Newhouse}}\ and\ \bibinfo {author} {\bibfnamefont {T.}~\bibnamefont
  {Pignataro}},\ }\bibfield  {title} {\bibinfo {title} {On the estimation of
  topological entropy},\ }\href@noop {} {\bibfield  {journal} {\bibinfo
  {journal} {Journal of statistical physics}\ }\textbf {\bibinfo {volume}
  {72}},\ \bibinfo {pages} {1331} (\bibinfo {year} {1993})}\BibitemShut
  {NoStop}%
\bibitem [{\citenamefont {Goodwyn}(1969)}]{goodwyn1969topological}%
  \BibitemOpen
  \bibfield  {author} {\bibinfo {author} {\bibfnamefont {L.~W.}\ \bibnamefont
  {Goodwyn}},\ }\bibfield  {title} {\bibinfo {title} {Topological entropy
  bounds measure-theoretic entropy},\ }\href@noop {} {\bibfield  {journal}
  {\bibinfo  {journal} {Proceedings of the American Mathematical Society}\
  }\textbf {\bibinfo {volume} {23}},\ \bibinfo {pages} {679} (\bibinfo {year}
  {1969})}\BibitemShut {NoStop}%
\bibitem [{\citenamefont {Bowen}(1973)}]{bowen1973topological}%
  \BibitemOpen
  \bibfield  {author} {\bibinfo {author} {\bibfnamefont {R.}~\bibnamefont
  {Bowen}},\ }\bibfield  {title} {\bibinfo {title} {Topological entropy for
  noncompact sets},\ }\href@noop {} {\bibfield  {journal} {\bibinfo  {journal}
  {Transactions of the American Mathematical Society}\ }\textbf {\bibinfo
  {volume} {184}},\ \bibinfo {pages} {125} (\bibinfo {year}
  {1973})}\BibitemShut {NoStop}%
\bibitem [{\citenamefont {Bowen}(1970)}]{bowen1970topological}%
  \BibitemOpen
  \bibfield  {author} {\bibinfo {author} {\bibfnamefont {R.}~\bibnamefont
  {Bowen}},\ }\bibfield  {title} {\bibinfo {title} {Topological entropy and
  axiom a},\ }in\ \href@noop {} {\emph {\bibinfo {booktitle} {Proc. Sympos.
  Pure Math}}},\ Vol.~\bibinfo {volume} {14}\ (\bibinfo {year} {1970})\ pp.\
  \bibinfo {pages} {23--41}\BibitemShut {NoStop}%
\bibitem [{\citenamefont {Westerweel}\ \emph {et~al.}(2013)\citenamefont
  {Westerweel}, \citenamefont {Elsinga},\ and\ \citenamefont
  {Adrian}}]{westerweel2013particle}%
  \BibitemOpen
  \bibfield  {author} {\bibinfo {author} {\bibfnamefont {J.}~\bibnamefont
  {Westerweel}}, \bibinfo {author} {\bibfnamefont {G.~E.}\ \bibnamefont
  {Elsinga}},\ and\ \bibinfo {author} {\bibfnamefont {R.~J.}\ \bibnamefont
  {Adrian}},\ }\bibfield  {title} {\bibinfo {title} {Particle image velocimetry
  for complex and turbulent flows},\ }\href@noop {} {\bibfield  {journal}
  {\bibinfo  {journal} {Annual Review of Fluid Mechanics}\ }\textbf {\bibinfo
  {volume} {45}},\ \bibinfo {pages} {409} (\bibinfo {year} {2013})}\BibitemShut
  {NoStop}%
\bibitem [{\citenamefont {Toschi}\ and\ \citenamefont
  {Bodenschatz}(2009)}]{toschi2009lagrangian}%
  \BibitemOpen
  \bibfield  {author} {\bibinfo {author} {\bibfnamefont {F.}~\bibnamefont
  {Toschi}}\ and\ \bibinfo {author} {\bibfnamefont {E.}~\bibnamefont
  {Bodenschatz}},\ }\bibfield  {title} {\bibinfo {title} {Lagrangian properties
  of particles in turbulence},\ }\href@noop {} {\bibfield  {journal} {\bibinfo
  {journal} {Annual review of fluid mechanics}\ }\textbf {\bibinfo {volume}
  {41}},\ \bibinfo {pages} {375} (\bibinfo {year} {2009})}\BibitemShut
  {NoStop}%
\bibitem [{\citenamefont {Schanz}\ \emph {et~al.}(2016)\citenamefont {Schanz},
  \citenamefont {Gesemann},\ and\ \citenamefont
  {Schr{\"o}der}}]{schanz2016shake}%
  \BibitemOpen
  \bibfield  {author} {\bibinfo {author} {\bibfnamefont {D.}~\bibnamefont
  {Schanz}}, \bibinfo {author} {\bibfnamefont {S.}~\bibnamefont {Gesemann}},\
  and\ \bibinfo {author} {\bibfnamefont {A.}~\bibnamefont {Schr{\"o}der}},\
  }\bibfield  {title} {\bibinfo {title} {Shake-the-box: Lagrangian particle
  tracking at high particle image densities},\ }\href@noop {} {\bibfield
  {journal} {\bibinfo  {journal} {Experiments in fluids}\ }\textbf {\bibinfo
  {volume} {57}},\ \bibinfo {pages} {1} (\bibinfo {year} {2016})}\BibitemShut
  {NoStop}%
\bibitem [{\citenamefont {Manoharan}\ \emph {et~al.}(2025)\citenamefont
  {Manoharan}, \citenamefont {Subramanian},\ and\ \citenamefont
  {Joy}}]{mdkl-nnq1}%
  \BibitemOpen
  \bibfield  {author} {\bibinfo {author} {\bibfnamefont {A.}~\bibnamefont
  {Manoharan}}, \bibinfo {author} {\bibfnamefont {S.}~\bibnamefont
  {Subramanian}},\ and\ \bibinfo {author} {\bibfnamefont {A.}~\bibnamefont
  {Joy}},\ }\bibfield  {title} {\bibinfo {title} {Topological entropy of
  stationary two-dimensional turbulence},\ }\href
  {https://doi.org/10.1103/mdkl-nnq1} {\bibfield  {journal} {\bibinfo
  {journal} {Phys. Rev. E}\ }\textbf {\bibinfo {volume} {112}},\ \bibinfo
  {pages} {015106} (\bibinfo {year} {2025})}\BibitemShut {NoStop}%
\bibitem [{\citenamefont {Stohl}(1998)}]{stohl1998computation}%
  \BibitemOpen
  \bibfield  {author} {\bibinfo {author} {\bibfnamefont {A.}~\bibnamefont
  {Stohl}},\ }\bibfield  {title} {\bibinfo {title} {Computation, accuracy and
  applications of trajectories—a review and bibliography},\ }\href@noop {}
  {\bibfield  {journal} {\bibinfo  {journal} {Atmospheric Environment}\
  }\textbf {\bibinfo {volume} {32}},\ \bibinfo {pages} {947} (\bibinfo {year}
  {1998})}\BibitemShut {NoStop}%
\bibitem [{\citenamefont {Peacock}\ and\ \citenamefont
  {Haller}(2013)}]{peacock2013lagrangian}%
  \BibitemOpen
  \bibfield  {author} {\bibinfo {author} {\bibfnamefont {T.}~\bibnamefont
  {Peacock}}\ and\ \bibinfo {author} {\bibfnamefont {G.}~\bibnamefont
  {Haller}},\ }\bibfield  {title} {\bibinfo {title} {Lagrangian coherent
  structures: The hidden skeleton of fluid flows},\ }\href@noop {} {\bibfield
  {journal} {\bibinfo  {journal} {Physics today}\ }\textbf {\bibinfo {volume}
  {66}},\ \bibinfo {pages} {41} (\bibinfo {year} {2013})}\BibitemShut {NoStop}%
\bibitem [{\citenamefont {Thiffeault}(2010)}]{thiffeault2010braids}%
  \BibitemOpen
  \bibfield  {author} {\bibinfo {author} {\bibfnamefont {J.-L.}\ \bibnamefont
  {Thiffeault}},\ }\bibfield  {title} {\bibinfo {title} {Braids of entangled
  particle trajectories},\ }\href@noop {} {\bibfield  {journal} {\bibinfo
  {journal} {Chaos: An Interdisciplinary Journal of Nonlinear Science}\
  }\textbf {\bibinfo {volume} {20}} (\bibinfo {year} {2010})}\BibitemShut
  {NoStop}%
\bibitem [{\citenamefont {Haszpra}\ and\ \citenamefont
  {T{\'e}l}(2013)}]{Hazpra-S}%
  \BibitemOpen
  \bibfield  {author} {\bibinfo {author} {\bibfnamefont {T.}~\bibnamefont
  {Haszpra}}\ and\ \bibinfo {author} {\bibfnamefont {T.}~\bibnamefont
  {T{\'e}l}},\ }\bibfield  {title} {\bibinfo {title} {Topological entropy: A
  lagrangian measure of the state of the free atmosphere},\ }\href
  {https://doi.org/10.1175/JAS-D-13-069.1} {\bibfield  {journal} {\bibinfo
  {journal} {Journal of the Atmospheric Sciences}\ }\textbf {\bibinfo {volume}
  {70}},\ \bibinfo {pages} {4030 } (\bibinfo {year} {2013})}\BibitemShut
  {NoStop}%
\bibitem [{\citenamefont {Haszpra}\ and\ \citenamefont
  {T{\'e}l}(2011)}]{haszpra2011volcanic}%
  \BibitemOpen
  \bibfield  {author} {\bibinfo {author} {\bibfnamefont {T.}~\bibnamefont
  {Haszpra}}\ and\ \bibinfo {author} {\bibfnamefont {T.}~\bibnamefont
  {T{\'e}l}},\ }\bibfield  {title} {\bibinfo {title} {Volcanic ash in the free
  atmosphere: A dynamical systems approach},\ }in\ \href@noop {} {\emph
  {\bibinfo {booktitle} {Journal of Physics: Conference Series}}},\ Vol.\
  \bibinfo {volume} {333}\ (\bibinfo {organization} {IOP Publishing},\ \bibinfo
  {year} {2011})\ p.\ \bibinfo {pages} {012008}\BibitemShut {NoStop}%
\bibitem [{\citenamefont {Ottino}\ \emph {et~al.}(1990)\citenamefont {Ottino}
  \emph {et~al.}}]{ottino1990mixing}%
  \BibitemOpen
  \bibfield  {author} {\bibinfo {author} {\bibfnamefont {J.~M.}\ \bibnamefont
  {Ottino}} \emph {et~al.},\ }\bibfield  {title} {\bibinfo {title} {Mixing,
  chaotic advection, and turbulence},\ }\href@noop {} {\bibfield  {journal}
  {\bibinfo  {journal} {Annual Review of Fluid Mechanics}\ }\textbf {\bibinfo
  {volume} {22}},\ \bibinfo {pages} {207} (\bibinfo {year} {1990})}\BibitemShut
  {NoStop}%
\bibitem [{\citenamefont {Zahtila}\ \emph {et~al.}(2023)\citenamefont
  {Zahtila}, \citenamefont {Chan}, \citenamefont {Ooi},\ and\ \citenamefont
  {Philip}}]{zahtila2023particle}%
  \BibitemOpen
  \bibfield  {author} {\bibinfo {author} {\bibfnamefont {T.}~\bibnamefont
  {Zahtila}}, \bibinfo {author} {\bibfnamefont {L.}~\bibnamefont {Chan}},
  \bibinfo {author} {\bibfnamefont {A.}~\bibnamefont {Ooi}},\ and\ \bibinfo
  {author} {\bibfnamefont {J.}~\bibnamefont {Philip}},\ }\bibfield  {title}
  {\bibinfo {title} {Particle transport in a turbulent pipe flow: direct
  numerical simulations, phenomenological modelling and physical mechanisms},\
  }\href@noop {} {\bibfield  {journal} {\bibinfo  {journal} {Journal of Fluid
  Mechanics}\ }\textbf {\bibinfo {volume} {957}},\ \bibinfo {pages} {A1}
  (\bibinfo {year} {2023})}\BibitemShut {NoStop}%
\bibitem [{\citenamefont {Kusters}(1991)}]{kusters1991influence}%
  \BibitemOpen
  \bibfield  {author} {\bibinfo {author} {\bibfnamefont {K.~A.}\ \bibnamefont
  {Kusters}},\ }\bibfield  {title} {\bibinfo {title} {The influence of
  turbulence on aggregation of small particles in agitated vessels},\
  }\href@noop {} {\  (\bibinfo {year} {1991})}\BibitemShut {NoStop}%
\bibitem [{\citenamefont {Ziemniak}\ \emph {et~al.}(1994)\citenamefont
  {Ziemniak}, \citenamefont {Jung},\ and\ \citenamefont
  {Tél}}]{ZIEMNIAK1994123}%
  \BibitemOpen
  \bibfield  {author} {\bibinfo {author} {\bibfnamefont {E.}~\bibnamefont
  {Ziemniak}}, \bibinfo {author} {\bibfnamefont {C.}~\bibnamefont {Jung}},\
  and\ \bibinfo {author} {\bibfnamefont {T.}~\bibnamefont {Tél}},\ }\bibfield
  {title} {\bibinfo {title} {Tracer dynamics in open hydrodynamical flows as
  chaotic scattering},\ }\href
  {https://doi.org/https://doi.org/10.1016/0167-2789(94)90255-0} {\bibfield
  {journal} {\bibinfo  {journal} {Physica D: Nonlinear Phenomena}\ }\textbf
  {\bibinfo {volume} {76}},\ \bibinfo {pages} {123} (\bibinfo {year}
  {1994})}\BibitemShut {NoStop}%
\bibitem [{\citenamefont {Eckmann}\ and\ \citenamefont
  {Ruelle}(1985)}]{RevModPhys.57.617}%
  \BibitemOpen
  \bibfield  {author} {\bibinfo {author} {\bibfnamefont {J.~P.}\ \bibnamefont
  {Eckmann}}\ and\ \bibinfo {author} {\bibfnamefont {D.}~\bibnamefont
  {Ruelle}},\ }\bibfield  {title} {\bibinfo {title} {Ergodic theory of chaos
  and strange attractors},\ }\href {https://doi.org/10.1103/RevModPhys.57.617}
  {\bibfield  {journal} {\bibinfo  {journal} {Rev. Mod. Phys.}\ }\textbf
  {\bibinfo {volume} {57}},\ \bibinfo {pages} {617} (\bibinfo {year}
  {1985})}\BibitemShut {NoStop}%
\bibitem [{\citenamefont {Zimmerman}\ \emph {et~al.}(2017)\citenamefont
  {Zimmerman}, \citenamefont {Morrill-Winter},\ and\ \citenamefont
  {Klewicki}}]{zimmerman2017design}%
  \BibitemOpen
  \bibfield  {author} {\bibinfo {author} {\bibfnamefont {S.}~\bibnamefont
  {Zimmerman}}, \bibinfo {author} {\bibfnamefont {C.}~\bibnamefont
  {Morrill-Winter}},\ and\ \bibinfo {author} {\bibfnamefont {J.}~\bibnamefont
  {Klewicki}},\ }\bibfield  {title} {\bibinfo {title} {Design and
  implementation of a hot-wire probe for simultaneous velocity and vorticity
  vector measurements in boundary layers},\ }\href@noop {} {\bibfield
  {journal} {\bibinfo  {journal} {Experiments in Fluids}\ }\textbf {\bibinfo
  {volume} {58}},\ \bibinfo {pages} {1} (\bibinfo {year} {2017})}\BibitemShut
  {NoStop}%
\bibitem [{\citenamefont {Cavo}\ \emph {et~al.}(2007)\citenamefont {Cavo},
  \citenamefont {Lemonis}, \citenamefont {Panidis},\ and\ \citenamefont
  {Papailiou}}]{cavo2007performance}%
  \BibitemOpen
  \bibfield  {author} {\bibinfo {author} {\bibfnamefont {A.}~\bibnamefont
  {Cavo}}, \bibinfo {author} {\bibfnamefont {G.}~\bibnamefont {Lemonis}},
  \bibinfo {author} {\bibfnamefont {T.}~\bibnamefont {Panidis}},\ and\ \bibinfo
  {author} {\bibfnamefont {D.}~\bibnamefont {Papailiou}},\ }\bibfield  {title}
  {\bibinfo {title} {Performance of a 12-sensor vorticity probe in the near
  field of a rectangular turbulent jet},\ }\href@noop {} {\bibfield  {journal}
  {\bibinfo  {journal} {Experiments in fluids}\ }\textbf {\bibinfo {volume}
  {43}},\ \bibinfo {pages} {17} (\bibinfo {year} {2007})}\BibitemShut {NoStop}%
\bibitem [{\citenamefont {Wallace}\ and\ \citenamefont
  {Vukoslav{\v{c}}evi{\'c}}(2010)}]{wallace2010measurement}%
  \BibitemOpen
  \bibfield  {author} {\bibinfo {author} {\bibfnamefont {J.~M.}\ \bibnamefont
  {Wallace}}\ and\ \bibinfo {author} {\bibfnamefont {P.~V.}\ \bibnamefont
  {Vukoslav{\v{c}}evi{\'c}}},\ }\bibfield  {title} {\bibinfo {title}
  {Measurement of the velocity gradient tensor in turbulent flows},\
  }\href@noop {} {\bibfield  {journal} {\bibinfo  {journal} {Annual review of
  fluid mechanics}\ }\textbf {\bibinfo {volume} {42}},\ \bibinfo {pages} {157}
  (\bibinfo {year} {2010})}\BibitemShut {NoStop}%
\bibitem [{\citenamefont {Mortensen}\ and\ \citenamefont
  {Langtangen}(2016)}]{mortensen2016high}%
  \BibitemOpen
  \bibfield  {author} {\bibinfo {author} {\bibfnamefont {M.}~\bibnamefont
  {Mortensen}}\ and\ \bibinfo {author} {\bibfnamefont {H.~P.}\ \bibnamefont
  {Langtangen}},\ }\bibfield  {title} {\bibinfo {title} {High performance
  python for direct numerical simulations of turbulent flows},\ }\href@noop {}
  {\bibfield  {journal} {\bibinfo  {journal} {Computer Physics Communications}\
  }\textbf {\bibinfo {volume} {203}},\ \bibinfo {pages} {53} (\bibinfo {year}
  {2016})}\BibitemShut {NoStop}%
\bibitem [{\citenamefont {Patterson}\ and\ \citenamefont
  {Orszag}(1971)}]{patterson1971spectral}%
  \BibitemOpen
  \bibfield  {author} {\bibinfo {author} {\bibfnamefont {G.}~\bibnamefont
  {Patterson}}\ and\ \bibinfo {author} {\bibfnamefont {S.~A.}\ \bibnamefont
  {Orszag}},\ }\bibfield  {title} {\bibinfo {title} {Spectral calculations of
  isotropic turbulence: Efficient removal of aliasing interactions},\
  }\href@noop {} {\bibfield  {journal} {\bibinfo  {journal} {Physics of
  Fluids}\ }\textbf {\bibinfo {volume} {14}},\ \bibinfo {pages} {2538}
  (\bibinfo {year} {1971})}\BibitemShut {NoStop}%
\bibitem [{\citenamefont {Alvelius}(1999)}]{alvelius1999random}%
  \BibitemOpen
  \bibfield  {author} {\bibinfo {author} {\bibfnamefont {K.}~\bibnamefont
  {Alvelius}},\ }\bibfield  {title} {\bibinfo {title} {Random forcing of
  three-dimensional homogeneous turbulence},\ }\href@noop {} {\bibfield
  {journal} {\bibinfo  {journal} {Physics of Fluids}\ }\textbf {\bibinfo
  {volume} {11}},\ \bibinfo {pages} {1880} (\bibinfo {year}
  {1999})}\BibitemShut {NoStop}%
\bibitem [{\citenamefont {Eswaran}\ and\ \citenamefont
  {Pope}(1988)}]{eswaran1988}%
  \BibitemOpen
  \bibfield  {author} {\bibinfo {author} {\bibfnamefont {V.}~\bibnamefont
  {Eswaran}}\ and\ \bibinfo {author} {\bibfnamefont {S.~B.}\ \bibnamefont
  {Pope}},\ }\bibfield  {title} {\bibinfo {title} {An examination of forcing in
  direct numerical simulations of turbulence},\ }\href@noop {} {\bibfield
  {journal} {\bibinfo  {journal} {Computers \& Fluids}\ }\textbf {\bibinfo
  {volume} {16}},\ \bibinfo {pages} {257} (\bibinfo {year} {1988})}\BibitemShut
  {NoStop}%
\bibitem [{\citenamefont {Chen}\ \emph {et~al.}(2015)\citenamefont {Chen},
  \citenamefont {Jin},\ and\ \citenamefont {Zhang}}]{chen2015lagrangian}%
  \BibitemOpen
  \bibfield  {author} {\bibinfo {author} {\bibfnamefont {J.-C.}\ \bibnamefont
  {Chen}}, \bibinfo {author} {\bibfnamefont {G.-D.}\ \bibnamefont {Jin}},\ and\
  \bibinfo {author} {\bibfnamefont {J.}~\bibnamefont {Zhang}},\ }\bibfield
  {title} {\bibinfo {title} {Lagrangian statistics in isotropic turbulent flows
  with deterministic and stochastic forcing schemes},\ }\href@noop {}
  {\bibfield  {journal} {\bibinfo  {journal} {Acta Mechanica Sinica}\ }\textbf
  {\bibinfo {volume} {31}},\ \bibinfo {pages} {25} (\bibinfo {year}
  {2015})}\BibitemShut {NoStop}%
\bibitem [{\citenamefont {Yeung}\ and\ \citenamefont
  {Pope}(1989)}]{yeung1989lagrangian}%
  \BibitemOpen
  \bibfield  {author} {\bibinfo {author} {\bibfnamefont {P.-K.}\ \bibnamefont
  {Yeung}}\ and\ \bibinfo {author} {\bibfnamefont {S.~B.}\ \bibnamefont
  {Pope}},\ }\bibfield  {title} {\bibinfo {title} {Lagrangian statistics from
  direct numerical simulations of isotropic turbulence},\ }\href@noop {}
  {\bibfield  {journal} {\bibinfo  {journal} {Journal of Fluid Mechanics}\
  }\textbf {\bibinfo {volume} {207}},\ \bibinfo {pages} {531} (\bibinfo {year}
  {1989})}\BibitemShut {NoStop}%
\bibitem [{\citenamefont {Crouter}\ and\ \citenamefont
  {Briens}(2019)}]{crouter2019methods}%
  \BibitemOpen
  \bibfield  {author} {\bibinfo {author} {\bibfnamefont {A.}~\bibnamefont
  {Crouter}}\ and\ \bibinfo {author} {\bibfnamefont {L.}~\bibnamefont
  {Briens}},\ }\bibfield  {title} {\bibinfo {title} {Methods to assess mixing
  of pharmaceutical powders},\ }\href@noop {} {\bibfield  {journal} {\bibinfo
  {journal} {AAPS PharmSciTech}\ }\textbf {\bibinfo {volume} {20}},\ \bibinfo
  {pages} {1} (\bibinfo {year} {2019})}\BibitemShut {NoStop}%
\bibitem [{\citenamefont {Ebrahimi}\ \emph {et~al.}(2019)\citenamefont
  {Ebrahimi}, \citenamefont {Tamer}, \citenamefont {Villegas}, \citenamefont
  {Chiappetta},\ and\ \citenamefont {Ein-Mozaffari}}]{ebrahimi2019application}%
  \BibitemOpen
  \bibfield  {author} {\bibinfo {author} {\bibfnamefont {M.}~\bibnamefont
  {Ebrahimi}}, \bibinfo {author} {\bibfnamefont {M.}~\bibnamefont {Tamer}},
  \bibinfo {author} {\bibfnamefont {R.~M.}\ \bibnamefont {Villegas}}, \bibinfo
  {author} {\bibfnamefont {A.}~\bibnamefont {Chiappetta}},\ and\ \bibinfo
  {author} {\bibfnamefont {F.}~\bibnamefont {Ein-Mozaffari}},\ }\bibfield
  {title} {\bibinfo {title} {Application of cfd to analyze the hydrodynamic
  behaviour of a bioreactor with a double impeller},\ }\href@noop {} {\bibfield
   {journal} {\bibinfo  {journal} {Processes}\ }\textbf {\bibinfo {volume}
  {7}},\ \bibinfo {pages} {694} (\bibinfo {year} {2019})}\BibitemShut {NoStop}%
\bibitem [{\citenamefont {Zoby}\ \emph {et~al.}(2011)\citenamefont {Zoby},
  \citenamefont {Navarro-Martinez}, \citenamefont {Kronenburg},\ and\
  \citenamefont {Marquis}}]{zoby2011turbulent}%
  \BibitemOpen
  \bibfield  {author} {\bibinfo {author} {\bibfnamefont {M.}~\bibnamefont
  {Zoby}}, \bibinfo {author} {\bibfnamefont {S.}~\bibnamefont
  {Navarro-Martinez}}, \bibinfo {author} {\bibfnamefont {A.}~\bibnamefont
  {Kronenburg}},\ and\ \bibinfo {author} {\bibfnamefont {A.}~\bibnamefont
  {Marquis}},\ }\bibfield  {title} {\bibinfo {title} {Turbulent mixing in
  three-dimensional droplet arrays},\ }\href@noop {} {\bibfield  {journal}
  {\bibinfo  {journal} {International journal of heat and fluid flow}\ }\textbf
  {\bibinfo {volume} {32}},\ \bibinfo {pages} {499} (\bibinfo {year}
  {2011})}\BibitemShut {NoStop}%
\bibitem [{\citenamefont {Li}\ \emph {et~al.}(2021)\citenamefont {Li},
  \citenamefont {Wang}, \citenamefont {Fang}, \citenamefont {Wang},
  \citenamefont {Liu}, \citenamefont {Tian}, \citenamefont {Qiu},\ and\
  \citenamefont {Su}}]{li2021cfd}%
  \BibitemOpen
  \bibfield  {author} {\bibinfo {author} {\bibfnamefont {J.}~\bibnamefont
  {Li}}, \bibinfo {author} {\bibfnamefont {M.}~\bibnamefont {Wang}}, \bibinfo
  {author} {\bibfnamefont {D.}~\bibnamefont {Fang}}, \bibinfo {author}
  {\bibfnamefont {J.}~\bibnamefont {Wang}}, \bibinfo {author} {\bibfnamefont
  {D.}~\bibnamefont {Liu}}, \bibinfo {author} {\bibfnamefont {W.}~\bibnamefont
  {Tian}}, \bibinfo {author} {\bibfnamefont {S.}~\bibnamefont {Qiu}},\ and\
  \bibinfo {author} {\bibfnamefont {G.}~\bibnamefont {Su}},\ }\bibfield
  {title} {\bibinfo {title} {Cfd simulation on the transient process of coolant
  mixing phenomenon in reactor pressure vessel},\ }\href@noop {} {\bibfield
  {journal} {\bibinfo  {journal} {Annals of Nuclear Energy}\ }\textbf {\bibinfo
  {volume} {153}},\ \bibinfo {pages} {108045} (\bibinfo {year}
  {2021})}\BibitemShut {NoStop}%
\bibitem [{\citenamefont {Biswas}\ \emph {et~al.}()\citenamefont {Biswas},
  \citenamefont {Jagannathan},\ and\ \citenamefont {Joy}}]{unpub-geophys-comm}%
  \BibitemOpen
  \bibfield  {author} {\bibinfo {author} {\bibfnamefont {A.}~\bibnamefont
  {Biswas}}, \bibinfo {author} {\bibfnamefont {A.}~\bibnamefont
  {Jagannathan}},\ and\ \bibinfo {author} {\bibfnamefont {A.}~\bibnamefont
  {Joy}},\ }\bibfield  {title} {\bibinfo {title} {Unpublished},\ }\href@noop {}
  {\ }\BibitemShut {NoStop}%
\bibitem [{\citenamefont {Srinivasan}\ \emph {et~al.}(2023)\citenamefont
  {Srinivasan}, \citenamefont {Barkan},\ and\ \citenamefont
  {McWilliams}}]{srinivasan2023forward}%
  \BibitemOpen
  \bibfield  {author} {\bibinfo {author} {\bibfnamefont {K.}~\bibnamefont
  {Srinivasan}}, \bibinfo {author} {\bibfnamefont {R.}~\bibnamefont {Barkan}},\
  and\ \bibinfo {author} {\bibfnamefont {J.~C.}\ \bibnamefont {McWilliams}},\
  }\bibfield  {title} {\bibinfo {title} {A forward energy flux at submesoscales
  driven by frontogenesis},\ }\href@noop {} {\bibfield  {journal} {\bibinfo
  {journal} {Journal of Physical Oceanography}\ }\textbf {\bibinfo {volume}
  {53}},\ \bibinfo {pages} {287} (\bibinfo {year} {2023})}\BibitemShut
  {NoStop}%
\bibitem [{\citenamefont {D’Asaro}\ \emph {et~al.}(2018)\citenamefont
  {D’Asaro}, \citenamefont {Shcherbina}, \citenamefont {Klymak},
  \citenamefont {Molemaker}, \citenamefont {Novelli}, \citenamefont {Guigand},
  \citenamefont {Haza}, \citenamefont {Haus}, \citenamefont {Ryan},
  \citenamefont {Jacobs} \emph {et~al.}}]{d2018ocean}%
  \BibitemOpen
  \bibfield  {author} {\bibinfo {author} {\bibfnamefont {E.~A.}\ \bibnamefont
  {D’Asaro}}, \bibinfo {author} {\bibfnamefont {A.~Y.}\ \bibnamefont
  {Shcherbina}}, \bibinfo {author} {\bibfnamefont {J.~M.}\ \bibnamefont
  {Klymak}}, \bibinfo {author} {\bibfnamefont {J.}~\bibnamefont {Molemaker}},
  \bibinfo {author} {\bibfnamefont {G.}~\bibnamefont {Novelli}}, \bibinfo
  {author} {\bibfnamefont {C.~M.}\ \bibnamefont {Guigand}}, \bibinfo {author}
  {\bibfnamefont {A.~C.}\ \bibnamefont {Haza}}, \bibinfo {author}
  {\bibfnamefont {B.~K.}\ \bibnamefont {Haus}}, \bibinfo {author}
  {\bibfnamefont {E.~H.}\ \bibnamefont {Ryan}}, \bibinfo {author}
  {\bibfnamefont {G.~A.}\ \bibnamefont {Jacobs}}, \emph {et~al.},\ }\bibfield
  {title} {\bibinfo {title} {Ocean convergence and the dispersion of flotsam},\
  }\href@noop {} {\bibfield  {journal} {\bibinfo  {journal} {Proceedings of the
  National Academy of Sciences}\ }\textbf {\bibinfo {volume} {115}},\ \bibinfo
  {pages} {1162} (\bibinfo {year} {2018})}\BibitemShut {NoStop}%
\bibitem [{\citenamefont {Jagannathan}\ \emph {et~al.}(2021)\citenamefont
  {Jagannathan}, \citenamefont {Srinivasan}, \citenamefont {McWilliams},
  \citenamefont {Molemaker},\ and\ \citenamefont
  {Stewart}}]{jagannathan2021boundary}%
  \BibitemOpen
  \bibfield  {author} {\bibinfo {author} {\bibfnamefont {A.}~\bibnamefont
  {Jagannathan}}, \bibinfo {author} {\bibfnamefont {K.}~\bibnamefont
  {Srinivasan}}, \bibinfo {author} {\bibfnamefont {J.~C.}\ \bibnamefont
  {McWilliams}}, \bibinfo {author} {\bibfnamefont {M.~J.}\ \bibnamefont
  {Molemaker}},\ and\ \bibinfo {author} {\bibfnamefont {A.~L.}\ \bibnamefont
  {Stewart}},\ }\bibfield  {title} {\bibinfo {title} {Boundary-layer-mediated
  vorticity generation in currents over sloping bathymetry},\ }\href@noop {}
  {\bibfield  {journal} {\bibinfo  {journal} {Journal of Physical
  Oceanography}\ }\textbf {\bibinfo {volume} {51}},\ \bibinfo {pages} {1757}
  (\bibinfo {year} {2021})}\BibitemShut {NoStop}%
\bibitem [{\citenamefont {Gradshteyn}\ and\ \citenamefont
  {Ryzhik}(2000)}]{GradshteynRyzhik2000}%
  \BibitemOpen
  \bibfield  {author} {\bibinfo {author} {\bibfnamefont {I.~S.}\ \bibnamefont
  {Gradshteyn}}\ and\ \bibinfo {author} {\bibfnamefont {I.~M.}\ \bibnamefont
  {Ryzhik}},\ }\href@noop {} {\emph {\bibinfo {title} {Tables of Integrals,
  Series, and Products}}},\ \bibinfo {edition} {6th}\ ed.\ (\bibinfo
  {publisher} {Academic Press},\ \bibinfo {address} {San Diego, CA},\ \bibinfo
  {year} {2000})\ \bibinfo {note} {see formula 4.291.8 for Serret's
  integral}\BibitemShut {NoStop}%
\end{thebibliography}%
\end{document}